\documentclass[twocolumn]{aastex63_cranmer2020}
\usepackage{times}

\shorttitle{Solar Wind Proton and Electron Heating Rates}
\shortauthors{S.\  R.\  Cranmer}

\begin{document}

\title{Heating Rates for Protons and Electrons in Polar Coronal Holes:
Empirical Constraints from the Ultraviolet Coronagraph Spectrometer}

\author[0000-0002-3699-3134]{Steven R. Cranmer}
\affiliation{Department of Astrophysical and Planetary Sciences,
Laboratory for Atmospheric and Space Physics,
University of Colorado, Boulder, CO 80309, USA}

\begin{abstract}
Ultraviolet spectroscopy of the extended solar corona is a powerful
tool for measuring the properties of protons, electrons, and heavy ions
in the accelerating solar wind.
The large coronal holes that expand up from the north and south poles
at solar minimum are low-density collisionless regions in which it is
possible to detect departures from one-fluid thermal equilibrium.
An accurate characterization of these departures is helpful
in identifying the kinetic processes ultimately responsible for
coronal heating.
In this paper, Ultraviolet Coronagraph Spectrometer (UVCS)
measurements of the H~I Lyman $\alpha$ line are analyzed to constrain
values for the solar wind speed, electron density, electron temperature,
proton temperature (parallel and perpendicular to the magnetic field)
and Alfv\'{e}n-wave amplitude.
The analysis procedure involves creating a large randomized ensemble
of empirical models, simulating their Ly$\alpha$ profiles, and building
posterior probability distributions for only the models that agree
with the UVCS data.
The resulting temperatures do not exhibit a great deal of radial
variation between heliocentric distances of 1.4 and 4 solar radii.
Typical values for the electron, parallel proton, and
perpendicular proton temperatures are 1.2, 1.8, and 1.9 MK,
respectively.
Resulting values for the ``nonthermal'' Alfv\'{e}n wave amplitude
show evidence for weak dissipation, with a total energy-loss rate
that agrees well with an independently derived total heating rate
for the protons and electrons.
The moderate Alfv\'{e}n-wave amplitudes appear to resolve some tension
in the literature between competing claims of both higher (undamped)
and lower (heavily damped) values.
\end{abstract}

\keywords{Alfv\'{e}n waves (23) --
Solar coronal holes (1484) --
Solar ultraviolet emission (1533) --
Solar wind (1534) --
Space plasmas (1544) --
Spectroscopy (1558)}

\section{Introduction}
\label{sec:intro}

The Sun's upper atmosphere is heated to temperatures greater than
$10^6$~K, and solar plasma flows out into the heliosphere at
supersonic speeds.
Even after almost a century of study, the physical processes
responsible for heating the corona and accelerating the solar
wind are still not understood
\citep[see, e.g.,][]{PD12,Kl15,Pe15,Ab16,CW19}.
In order to test any theoretical model that proposes a new physical
process, there must be accurate and relevant observational data.
Thus, when investigating the origin of the solar wind, we need to know
particle speeds, densities, and temperatures at heliocentric distances
between about 1.5 and 5 solar radii ($R_{\odot}$).
This is the ``extended corona'' (also called the middle corona) 
where much of the solar wind's acceleration occurs.
It is also the region in which the corona begins to evolve from being
a collisional magnetohydrodynamic (MHD) fluid to being a
collisionless kinetic plasma.

Observations of the extended corona are somewhat rare because this
region sits in a gap between the distances probed by standard solar
telescopes and the distances probed by many space-based coronagraphs.
Fortunately, there have been instruments designed specifically
to measure plasma properties in this gap.
This paper focuses on observations made by the
Ultraviolet Coronagraph Spectrometer (UVCS), one of twelve
instruments on the {\em Solar and Heliospheric Observatory}
({\em{SOHO}}) spacecraft \citep{Do95,FS97}.
From 1996 to 2013, UVCS observed emission lines with wavelengths
between 470 and 1360~{\AA} in the extended corona
\citep{Ko95,Ko97,Ko06a,No97,An06}.

In the collisionless extended corona, different populations of
particles (e.g., neutral atoms, protons, electrons, and heavy ions)
can exhibit different thermodynamic properties.
When comparing any two populations, there can be differential flows,
unequal temperatures, and differing amounts of anisotropic departure
from an equilibrium Maxwell-Boltzmann velocity distribution
\citep[see][]{Ma06,Ve19}.
UVCS has been used to measure some strong departures from thermal
equilibrium for heavy ions such as O$^{+5}$ and Mg$^{+9}$
\citep[e.g.,][]{Ko99,Cr08}.
These observations have been incredibly useful as diagnostics of
collisionless particle energization.
However, heavy ions contribute only a tiny fraction of the total
mass, momentum, and energy of the corona.

To learn more about the processes that heat and accelerate the
corona as a whole, one needs to study hydrogen, the dominant
elemental constituent.
At coronal temperatures of $10^6$~K, hydrogen is almost fully
ionized, so a complete kinetic description requires knowledge of
the properties of both protons and electrons.
The small remaining population of neutral hydrogen atoms enables
the formation of spectral-line photons, and in this paper we analyze
UVCS observations of the bright H~I Ly$\alpha$ (1215.67~{\AA})
resonance line.
Because the neutrals are closely coupled to the protons by
collisions and charge-exchange processes, the properties of the
H~I Ly$\alpha$ line are sensitive to {\em proton flow speeds and
anisotropic temperatures.}
Also, the ionization/recombination balance that determines the
time-steady neutral hydrogen concentration depends on the
{\em electron temperature} \citep[see][]{Gb71,Ko80,Wi82,St93}.

The goal of this paper is to show how UVCS H~I Ly$\alpha$ data
can be combined with some well-established physical principles
(e.g., mass and momentum conservation) to provide self-consistent
measurements of proton and electron temperatures, outflow speeds,
and ``nonthermal'' wave/turbulence amplitudes.
Section \ref{sec:uvcs} presents the UVCS data, which was taken
from north and south polar coronal holes during the 1996--1997
solar minimum.
Section \ref{sec:emp} describes the construction of a large-scale
Monte Carlo ensemble of trial ``empirical models'' that are meant
to explore the full parameter space of possible proton and electron
properties.
Section \ref{sec:lya} then summarizes how we synthesize a set of
predicted H~I Ly$\alpha$ line profiles for each of the trial
models, and Section \ref{sec:results} presents the comparison
between the synthetic and observed profiles.
By taking only the small subset of models that agree with the
UVCS data, we produce a set of posterior probability distributions
for the relevant proton and electron properties.
Section \ref{sec:conc} concludes by summarizing some broader
implications of this work and suggesting future improvements.
The Appendix contains a supplementary estimation of the
proton and electron heating rates that appear to be required to
maintain the corona in its measured state.

\section{UVCS Observations}
\label{sec:uvcs}

This paper is concerned with the plasma properties of large
polar coronal holes that are known to be associated with high-speed
solar wind streams \citep[see, e.g.,][]{HS79,Wa09}.
For a few years around each solar minimum, the Sun's magnetic field
appears to expand superradially above its north and south poles,
and there is a roughly axisymmetric magnetic geometry.
The UVCS data analyzed in this paper consist of observations over the
poles during the 1996--1997 minimum between Solar Cycles 22 and 23.

The UVCS instrument contains two ultraviolet toric-grating spectrometers
paired with a system of external and internal occulters that block out
light from the bright solar disk.
The spectrometer slits can be rotated around the Sun and are always
oriented tangentially to the limb.
Each of the observations discussed below was performed at a nominal
heliocentric distance $r$, defined by the intersection between the slit
and a line extending out radially from the Sun, between 1.4 and 4.1
solar radii ($R_{\odot}$).
The total length of the slit corresponds to 40$'$ on the sky (i.e.,
2.5 $R_{\odot}$ in the corona), and the width of the slit is adjustable
depending on the desired count rate and spectral resolution.

The observational data used in this paper are the intensities and
widths of coronal H~I Ly$\alpha$ emission lines.
The specific UVCS measurements are the same as those presented by
\citet{Cr99}.
Although other UVCS data exist for the 1996--1997 solar minimum
\citep[see, e.g.,][]{Ko97,Es99,Su99,Zn99,An00,Nk08,St12,Do16},
the selected data represent a homogeneous collection of observations
that was reduced and analyzed with identical procedures.
Specifically, this paper excludes all data taken when part or all
of the UVCS slit was exposed directly to the solar disk.
Observations in that mode have not been as well-calibrated as the
more standard off-limb, over-occulted mode.

As described by \citet{Cr99}, the H~I Ly$\alpha$ data were obtained
mainly in two time periods.
The first was 1996 December 28 to 1997 January 5, corresponding to
Carrington Rotation (CR) 1917, with UVCS lead scientist S.\  Fineschi.
The second was 1997 April 14 to 1997 April 20, corresponding to CR 1921,
with UVCS lead scientist S.\  Cranmer.
Additional context on the nearly axisymmetric structure of the corona
at solar minimum was obtained from the first ``Whole Sun Month''
campaign in CR 1913 \citep{GK99,Gu99}.

\citet{Cr99} described how the UVCS data analysis software (DAS)
was used to compensate for image distortion, correct for detector
flat-field effects, and calibrate the data in intensity and wavelength
\citep[see also][]{Ga96,Ga02,Ko06a}.
The total emission at H~I Ly$\alpha$ wavelengths was corrected for
instrument-scattered stray light from the solar disk \citep{Cr10}
and emission due to interstellar neutral hydrogen atoms in the outer
heliosphere \citep[e.g.,][]{Bx97,Sp17}.
An additional correction to the heliocentric slit distance has been
implemented to account for cross-talk between the UVCS
mirror-pointing mechanism and the grating mechanism.
For most of the H~I Ly$\alpha$ data, this amounted to no more than a
3\% relative correction to the ``commanded'' radial distance;
for $r \lesssim 1.6 \, R_{\odot}$ it was more like a 6\% correction.
To obtain the individual intensity measurements discussed below,
counts were summed over the central 15$'$--25$'$ of the UVCS
spectrometer slit, which averages over the differences between
high-density polar plumes and low-density interplume regions.

\begin{figure}[!t]
\epsscale{1.17}
\plotone{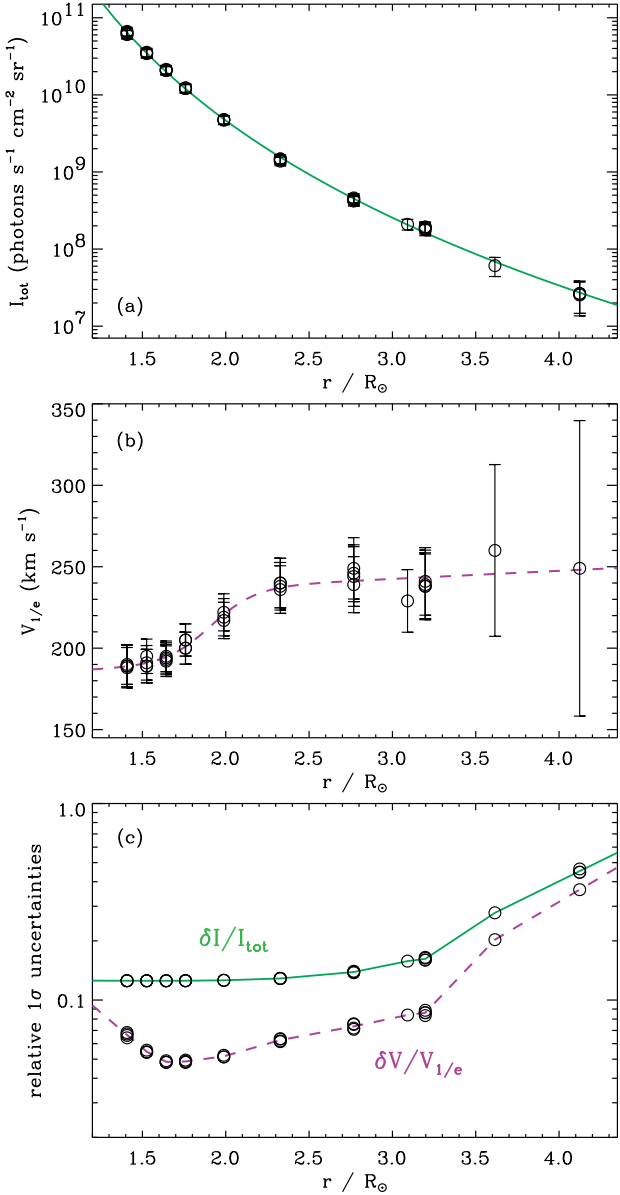}
\caption{H~I Ly$\alpha$ coronal hole measurements from
\citet{Cr99}:
(a) total line-integrated intensities, with fit function from
Equation (\ref{eq:Itotfit}),
(b) $1/e$ half-widths expressed in Doppler velocity units, with
fit function from Equation (\ref{eq:V1efit}),
and (c) ratios of 1$\sigma$ uncertainties to associated means,
with curves showing linearly interpolated values.
\label{fig01}}
\end{figure}

Figure \ref{fig01} shows the calibrated H~I Ly$\alpha$ intensities
(integrated over the coronal emission line profile) and their
$1/e$ Gaussian half-widths $\Delta \lambda_{1/e}$
as a function of radial distance at slit-center.
The half-widths are shown in Doppler velocity units as
$V_{1/e} = c \Delta \lambda_{1/e} / \lambda_0$, where $c$ is the
speed of light and $\lambda_0$ is the rest wavelength of the line.
Error bars provide $\pm 1 \sigma$ uncertainty limits that were
computed from Poisson count-rate statistics and known
uncertainties in the various instrumental correction steps
\citep[see][]{Cr99}.
Because the data were obtained with somewhat large gaps between
successive radii, it is useful to provide smooth fitting functions
so that values at intermediate radii can be estimated.
For the total intensity,
\begin{equation}
  I_{\rm tot} \, = \, I_0 \left( \frac{5.092}{x^{10.04}} +
  \frac{5.238}{x^{6.972}} \right) \,\, ,
  \label{eq:Itotfit}
\end{equation}
where $x = r / R_{\odot}$ and
$I_0 = 10^{11}$ photons s$^{-1}$ cm$^{-2}$ sr$^{-1}$.
For the Gaussian half-width,
\begin{equation}
  V_{1/e} \, = \, A \, + \, B x \, + \, C \tanh
  \left( \frac{x - 1.9107}{0.27154} \right)
  \label{eq:V1efit}
\end{equation}
where $A = 204.21$ km~s$^{-1}$, $B = 4.9612$ km~s$^{-1}$, and
$C = 23.433$ km~s$^{-1}$.
Figure \ref{fig01}(c) shows the radial dependence of the fractional
uncertainties of both quantities.
Instead of fitting these to smooth functions, it is straightforward
to use linear interpolation (using the mean values at each discrete
radial pointing) to specify their values at intermediate radii.

\section{The Ensemble of Empirical Models}
\label{sec:emp}

The observed properties of the H~I Ly$\alpha$ emission line depend
on a combination of several parameters associated with hydrogen and
free electrons in the extended corona.
The goal of this paper is to put realistic constraints on values for
as many of these particle parameters as possible.
In general, it is not straightforward to ``invert the data;'' i.e.,
to compute the desired parameters directly from the intensities and
line widths.
Instead, an empirical {\em forward modeling} approach is adopted.
In this approach, a large Monte Carlo ensemble of randomized trials is
created for the particle parameters.
Synthetic H~I Ly$\alpha$ profiles are simulated for each trial model.
Comparison with the actual observed data provides a way to select which
sets of particle parameters are the most realistic.

\begin{table*}[!ht]
\begin{center}
\caption{Monte Carlo Parameter Ranges}
\label{tab01}
\begin{tabular}{llcc}
\hline
\hline
\multicolumn{2}{c}{Quantity} & Minimum & Maximum \\
\hline
$\Delta_{\rm lo}$ &
Lower electron density multiplier & 1 & 4 \\
$\Delta_{\rm hi}$ &
Upper electron density multiplier & 1 & 4 \\
$u_{\infty}$ &
Wind speed at 1 AU (km~s$^{-1}$) & 350 & 900 \\
$f_{\rm max}$ &
Superradial expansion factor & 6.8 & 10.2 \\
$\xi_0$ &
Turbulent broadening at $r = 1.4 \, R_{\odot}$ (km~s$^{-1}$) & 30 & 90 \\
$\Psi$ &
Proton perpendicular speed multiplier & 0.7 & 1.3 \\
$\alpha$ &
Proton temperature anisotropy ratio & 0.1 & 10 \\
$T_{\rm top}$ &
Electron temperature at $r = 5 \, R_{\odot}$ (MK) & 0.3 & 3 \\
\hline
\end{tabular}
\end{center}
\end{table*}

The following subsections describe the Monte Carlo model parameters,
and Table \ref{tab01} lists them along with their allowed ranges.
Specifically, each trial empirical model is described by specific
choices for the 8 parameters listed in Table~\ref{tab01}.
Once those parameters are chosen, the procedures described in
Sections \ref{sec:emp:ne}--\ref{sec:emp:Te} specify the descriptions
of continuous radial functions for the electron density $n_e$, 
solar wind proton speed $u_p$,
radial magnetic field magnitude $B_r$,
Alfv\'{e}n speed $V_{\rm A}$,
Alfv\'{e}n wave velocity amplitude $v_{\perp}$,
bi-Maxwellian proton temperature components $T_{p \parallel}$ and
$T_{p \perp}$, and isotropic electron temperature $T_e$.
Since the goal of this paper is to model the H~I Ly$\alpha$ data
from the 1996--1997 solar minimum, the other observational constraints
on the Monte Carlo parameters should come from this time period, too.

In earlier studies of UVCS coronal-hole data \citep[e.g.,][]{Cr99,Cr08}
the approach was to never presume that the empirical plasma properties
must obey any specific physical laws.
In other words, it may prejudice the result if a specific assumption
about, say, the coronal heating mechanism was included in the analysis
that leads to a measured value of the coronal temperature.
This ``empirical modeling'' is meant to be clearly distinct
from the bottom-up kind of modeling that involves, say, solving the
full set of MHD conservation equations in a self-consistent way.
However, there are some nearly universal physical laws that we
expect to be valid in the extended corona for long time-averages,
such as the steady-state conservation of mass and linear momentum
along magnetic field lines.
Thus, in a slight departure from earlier empirical
modeling work (i.e., following the example of
\citeauthor{LS16} \citeyear{LS16}), these laws are utilized
in this paper to help put firmer constraints on the time-steady
plasma properties.
Nevertheless, we continue to refrain from making any assumptions about
the physical processes directly responsible for heating the corona.

\subsection{Coronal Number Densities}
\label{sec:emp:ne}

\begin{figure}[!t]
\epsscale{1.17}
\plotone{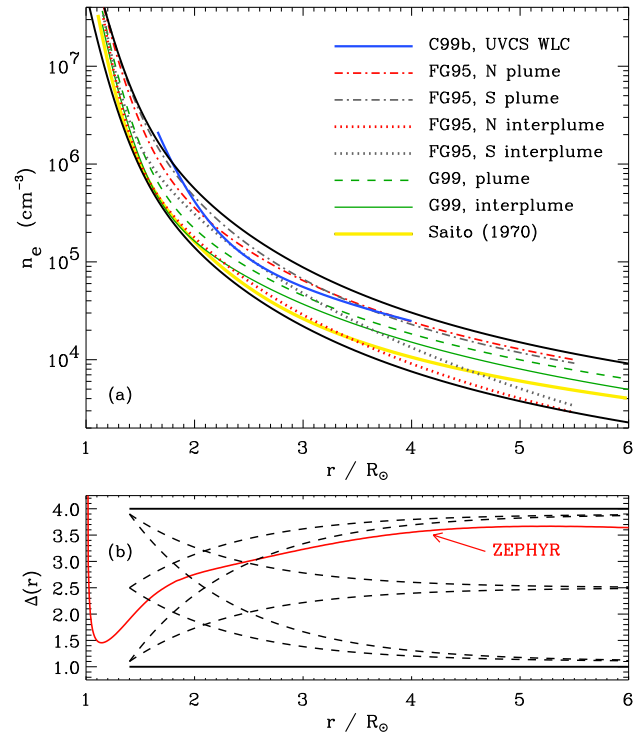}
\caption{(a) Radial dependence of electron density from
\citet{Cr99} (labeled C99b), \citet{FG95} (labeled FG95), and
\citet{Gu99} (labeled G99), and \citet{Saito}.
The lower black curve is from Equation (\ref{eq:ne0}) and the
upper black curve is from Equation (\ref{eq:nefull}) with
$\Delta_{\rm lo} = \Delta_{\rm hi} = 4$.
(b) Example parameter ranges for $\Delta(r)$ (black curves),
compared with a theoretical model curve obtained from solving
Equation (\ref{eq:nefull}) for $\Delta$ (red curve).
\label{fig02}}
\end{figure}

For the coronal regions modeled in this paper, the plasma is
believed to be almost fully ionized.
Thus, measurements of the free-electron number density $n_e$ provide
a reliable proxy for the total plasma density.
There has been almost a century's worth of electron-density measurement
using the linearly polarized component of the Thomson-scattered visible
continuum above the solar limb \citep{Min30,vd50,In15}.
Figure \ref{fig02}(a) shows several radial $n_e$ curves derived from
these kind of observations in polar coronal holes.
The measurements were obtained from space-based instruments such as the
{\em Spartan 201} White Light Coronagraph \citep{FG95},
the Large Angle Spectroscopic Coronagraph (LASCO) on {\em SOHO}
\citep{Gu99}, and the UVCS White Light Channel \citep[WLC;][]{Cr99}.
Additional data from near the solar limb were also obtained from
the Mauna Loa Mark III K-coronameter \citep{Gu99}.
For context we also show the electron density reconstruction
of \citet{Saito}, which was found to match observations over the north
and south heliographic poles at earlier solar-minimum phases.

Taking into account their varying absolute normalizations, the
electron density curves in Figure \ref{fig02}(a) appear to have
rather similar radial shapes.
However, there are subtle differences between them; see also
Figure~3 of \citet{Cr08}.
These differences were found to be important when evaluating radial
pressure-gradient forces (Section \ref{sec:emp:Te}),
so they must be modeled carefully.
Thus, we employ several free parameters to specify the radial shape
of $n_e(r)$ in the Monte Carlo ensemble models.
To start, it is noted that the electron density never seems to
dip below a minimum ``floor'' function, which is given by
\begin{equation}
  n_{e,0} \, = \, 10^{5} \,\, \mbox{cm}^{-3} 
  \left( \frac{0.657}{x^2} + \frac{35.4}{x^{5}} +
  \frac{536}{x^{12}} \right)
  \, ,
  \label{eq:ne0}
\end{equation}
where $x = r / R_{\odot}$.
This functional form is similar to those used by
\citet{Doy99}, \citet{CvB05}, \citet{SC20}, and others.
The above expression is valid only for $x \geq 1.4$.
At lower heights, there need to be additional terms with higher
inverse powers of $x$ to account for the smaller density scale height
near the solar surface.

The data shown in Figure \ref{fig02}(a) indicate an
approximate dynamic range of about a factor of four
between the different measurements, so the complete
model expression for electron density is specified as
\begin{equation}
  n_e (r) \, = \, n_{e,0}(r) \, \Delta(r) \,\,\, ,
  \label{eq:nefull}
\end{equation}
where $\Delta(r)$ is a dimensionless multiplier that ranges
in value between 1 and 4.
Its radial dependence is given by
\begin{equation}
  \Delta(r) \, = \, \Delta_{\rm hi} +
  (\Delta_{\rm lo} - \Delta_{\rm hi}) e^{1.4-x}
\end{equation}
which is applied only for $x \geq 1.4$.
Figure \ref{fig02}(b) shows a selection of $\Delta(r)$ curves
for a coarse grid of values for the dimensionless parameters
$\Delta_{\rm lo}$ and $\Delta_{\rm hi}$,
each of which is also varied between values of 1 and 4.
Also shown is a scaled $\Delta (r)$ function taken from a
theoretical ZEPHYR model of a polar coronal hole \citep{CvB07}.

Specific choices for $\Delta_{\rm lo}$ and $\Delta_{\rm hi}$
are able to reproduce most of the observational curves shown
in Figure \ref{fig02}(a).
One possible exception is the $n_e$ curve derived from the
WLC instrument of UVCS.
The inferred density values at low heights ($r < 2 \, R_{\odot}$)
fall above the envelope of Monte Carlo model parameters, and
\citet{Cr99} discussed some possible instrumental effects that may
have contaminated these data.
Thus, we do not believe it is important to raise the maximum
values of $\Delta_{\rm lo}$ and $\Delta_{\rm hi}$ in order to
accommodate that particular observation.

Particle densities at 1~AU are measured frequently by in~situ
instruments in the solar wind, so it is worthwhile to test the
empirical model by comparing with these measurements.
The densities in interplanetary space are insensitive to the values
chosen for $\Delta_{\rm lo}$.
Given that $n_{e,0}(r)$ is fixed, the value of $n_e$ at
1~AU has a linear, one-to-one relationship with the value of
$\Delta_{\rm hi}$.
For a range of values for $\Delta_{\rm hi}$ between 1 and 4,
this model produces a range of values for $n_e$ at
1~AU between 1.42 and 5.68 cm$^{-3}$.
In comparison, proton density data from the OMNI database \citep{KP05}
was collected for the 25-year period between 1990 and 2015 and
converted to $n_e$ as discussed below.
Taking only fast solar wind data (bulk speeds greater than
600 km~s$^{-1}$), the median value of $n_e$ was found to be
3.08 cm$^{-3}$, and the distribution of values ranged between
1.65 and 6.05 cm$^{-3}$ for the 10\% and 90\% percentiles,
respectively.
This represents a dynamic range of 3.67, close to the value of 4
used to specify $\Delta_{\rm lo}$ and $\Delta_{\rm hi}$.

Models for the H~I Ly$\alpha$ emission require specifying the hydrogen
number density, and some of the conservation laws used below
require specifying the total mass density $\rho$.
Strictly speaking, the total hydrogen number density is given by
$n_{\rm H} = n_{\rm HI} + n_p$, where $n_{\rm HI}$ is the neutral
hydrogen number density and $n_p$ is the proton density.
However, at the coronal temperatures examined in this paper, it is
always the case that $n_{\rm HI} \ll n_p$.
Thus, both $n_p$ and $\rho$ can be computed under the approximation
of complete ionization, with
\begin{equation}
  n_p \, = \, \frac{n_e}{1 + 2h}
  \,\,\,\, \mbox{and} \,\,\,\,\,
  \rho \, = \, (1 + 4h) n_p m_p \,\, ,
\end{equation}
where $m_p$ is the proton mass and $h$ is the helium-to-hydrogen
number-density ratio, which is set to a typical value of 0.05
\citep[see, e.g.,][]{Ka07}.
The neutral hydrogen density is computed under the assumption of
local coronal ionization equilibrium, in which the local ratio 
$n_{\rm HI}/n_p$ is a function of the electron temperature $T_e$ only.
To specify this ratio, equilibrium tables from the CHIANTI
version 7.1 atomic database were used \citep{Dr97,Ln13},
and these results span a wide range of $T_e$ values between
$10^4$ and $10^9$ K.
Although the codes described below interpolate values from this table,
it is also useful to provide an approximate analytic expression.
Between $T_e$ values of about $10^6$ and $10^8$ K, the equilibrium
CHIANTI curve can be fit to about 10\% accuracy by
\begin{equation}
  \frac{n_{\rm HI}}{n_p} \, \approx \, \left(
  \frac{0.59}{T_e} \right)^{1.063} \,\, ,
\end{equation}
where $T_e$ is expressed in K.
At large heights in the low-density solar wind, coronal ionization
equilibrium is not always a valid assumption.
\citet{Sp17} showed that collisionless departures from ionization
equilibrium are not important for neutral hydrogen at most of the
coronal heights considered in this paper.
These effects begin to affect local values of $n_{\rm HI}/n_p$ 
substantially (i.e., at a $\sim$20\% level) above heights of
$r \approx 4 \, R_{\odot}$, and are much less important at lower heights.

\subsection{Bulk Solar Wind Acceleration}
\label{sec:emp:up}

The H~I Ly$\alpha$ emission is sensitive to the bulk velocity of
outflowing solar wind plasma.
In this paper, it is presumed that the different particle components
of the plasma (i.e., electrons, protons, and neutral hydrogen atoms)
are all flowing out at the same speed $u_p(r)$.\footnote{%
Differential flows between particle species tend to set in at even
larger heights than departures from collisional ionization equilibrium
\citep[e.g.,][]{Al00}, so they are assumed to be negligibly small
for the empirical models of the corona described here.}
Time-steady mass-flux conservation demands that the product
$\rho u_p A$ remains constant, where $A$ is the cross-sectional
area of a bundle of closely-spaced magnetic field lines, commonly
referred to as a flux tube.
Open flux tubes in the corona are often specified with an area
given by $A(r) = r^2 f(r)$, where $f(r)$ is a
superradial expansion factor normalized to $f=1$ at $r = R_{\odot}$.
Thus, anchoring the mass-flux normalization to its value at 1~AU, one
can solve for the hydrogen outflow speed at an arbitrary radius as
\begin{equation}
  u_p \, = \, u_{\infty} \,
  \frac{( n_e \, r^2 )_{\rm 1AU} \, f_{\rm max}}{n_e \, r^2 \, f}
  \,\, ,
\end{equation}
where $u_{\infty}$ and $f_{\rm max}$ are the values of $u_p$ and $f$
at 1~AU, respectively.

\begin{figure}[!t]
\epsscale{1.17}
\plotone{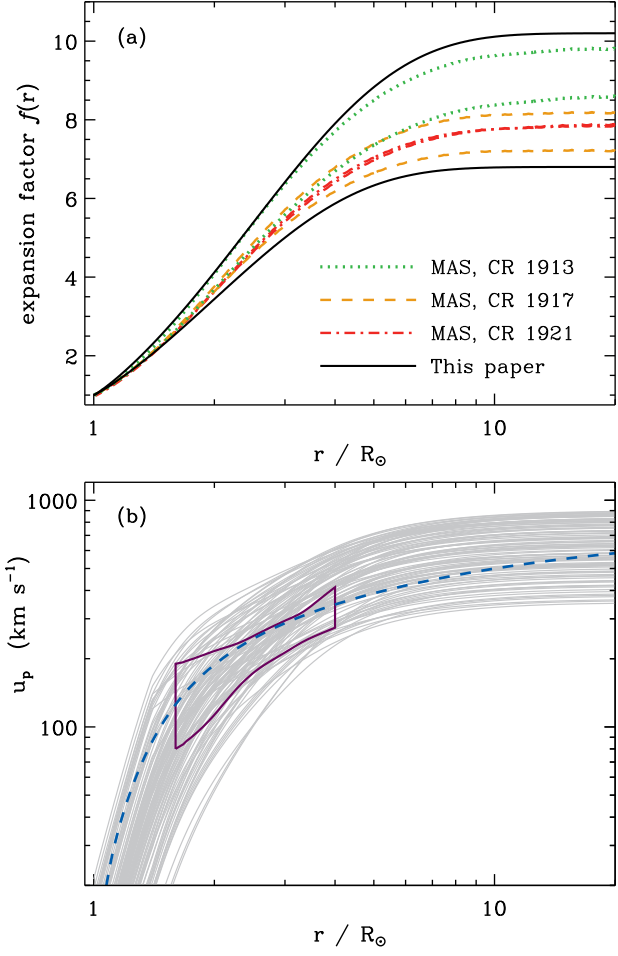}
\caption{(a) Superradial expansion factors for the MAS simulations
described in the text.  Each model corresponds to two curves: one for
the north polar field line, and one for the south polar field line.
(b) Modeled outflow speed profiles for 150 random trials
(gray solid curves).
Also shown is the theoretical ZEPHYR polar outflow speed
\citep[][blue dashed curve]{CvB07} and an earlier empirical estimate
of proton outflow speed from UVCS data \citep[][violet outline]{Cr99}.
\label{fig03}}
\end{figure}

Superradial expansion factors were computed for solar wind flows
over the north and south heliographic poles by examining output
from three-dimensional solutions of the polytropic MHD conservation
equations, constrained by synoptic magnetograms.
These solutions were computed by the Magnetohydrodynamics Around a
Sphere (MAS) code \citep[e.g.,][]{Mk99,Li99,Ri01,Ri07}
and made available by the MHDweb project.\footnote{%
http://www.predsci.com/mhdweb/}
Models were obtained for
CR 1913, 1917, and 1921, which correspond to the relevant times
for the UVCS observations (see Section \ref{sec:uvcs}).
The models for CR 1913 used photospheric magnetograms from the
National Solar Observatory (NSO) Kitt Peak Vacuum Telescope
\citep{Li76,Jn92} as a lower boundary condition.
The models for CR 1917 and 1921 used magnetgrams from the
{\em SOHO} Michelson Doppler Imager \citep[MDI;][]{Sc95}.
Figure \ref{fig03}(a) shows the polar $f(r)$ curves for these MAS models.
To parameterize these models in the Monte Carlo empirical-model
framework of this paper, a fitting function was found that reproduces
these numerical curves with only one free parameter ($f_{\rm max}$).
This function is given by
\begin{equation}
  f(r) \, = \, f_{\rm max} \, + \, (1 - f_{\rm max})
  \exp \left[ - \left( \frac{x-1}{H} \right)^{1.1} \right] \, ,
  \label{eq:fsuper}
\end{equation}
where $H=(f_{\rm max}/2.804)^{0.62}$ and
$x=r/R_{\odot}$ as above \citep[see also][]{SC20}.
The minimum and maximum values of $f_{\rm max}$ shown in 
Figure \ref{fig03}(a) are 6.8 and 10.2.
This range of values encompasses many earlier estimates of the
superradial expansion rate over the poles at solar minimum
\citep[see, e.g.,][]{KH76,MJ77,Ba98}.

With the density and superradial expansion factor both specified
with choices for the Monte Carlo parameters
$\Delta_{\rm lo}$, $\Delta_{\rm hi}$, and $f_{\rm max}$,
one needs only to choose a value for the parameter $u_{\infty}$
in order to specify the outflow speed as a function of radial distance.
Figure \ref{fig03}(b) shows the outflow speeds that result from 150
sets of randomized choices of these four parameters.
Despite the knowledge that polar coronal holes tend to contain
fast solar wind, we judged it to be too restrictive to rule
out the possibility that UVCS data may instead be consistent with
slow wind speeds in the corona.
Thus, trial values of $u_{\infty}$ were sampled from a uniform
probability distribution that ranges from 350 to 900 km~s$^{-1}$.
As seen in Figure \ref{fig03}(b), this range of values is consistent
with both theoretical models and earlier empirical determinations of
the wind speed from UVCS data.

\subsection{Magnetic Field and MHD Turbulence}
\label{sec:emp:Brxi}

Although the coronal magnetic field strength is not needed directly for
computing the H~I Ly$\alpha$ emission (at least for the unpolarized
component of the line observed by UVCS), it is needed to model the
MHD fluctuations that presumably broaden the line profiles
and affect momentum conservation.
Thus, the Monte Carlo empirical models specify a radial dependence
for the radial component of the magnetic field over the poles,
as well as the Alfv\'{e}n speed,
\begin{equation}
  V_{\rm A} \, = \, \frac{B_r}{\sqrt{4\pi\rho}} \,\, .
\end{equation}
The relative radial variation of $B_r$ is known already from the
superradial expansion (i.e., flux conservation demands
$B_r \propto A^{-1}$), but its absolute normalization remains to
be determined.
A useful common distance at which to set this normalization is 1~AU,
and we chose to examine the same OMNI database described above.
Using the same cutoff for fast wind speeds greater than 600
km~s$^{-1}$, the median value of the
radial field strength at 1~AU, which we call $B_{r \infty}$,
is 3.1 nT.
The associated 10\% and 90\% percentile values in this parameter
were 0.9 and 6.1 nT, respectively.
After some initial trial models in which $B_{r \infty}$ was sampled
randomly from this range, it was found that the results do
not depend strongly on this value.
Thus, to reduce the multidimensional space to be explored by the
Monte Carlo models, the decision was made to fix the value of
$B_{r \infty}$ to the median value of 3.1 nT for all models.

Given the above description of how the radial dependences of both
$B_r$ and $\rho$ are computed, a distribution of likely $V_{\rm A}$
profiles can be derived as well.
This distribution consists of curves that all have a single global
maximum in the extended corona, typically at values of $r$ between
1.5 and 1.8 $R_{\odot}$.
The maxima of $V_{\rm A}$ are distributed with a median value of
3105 km~s$^{-1}$ and approximate 10\% and 90\% percentile values of
2540 and 4140 km~s$^{-1}$, respectively.
The radial dependence of $V_{\rm A}$ is needed in order to describe
the spatial evolution of small-amplitude MHD fluctuations.
The time-steady damped wave-action-conservation model used here is
the same as the one described in more detail by \citet{CvB12}, with
\begin{equation}
  \frac{\partial}{\partial r} \left[
  \frac{(u_p + V_{\rm A})^2 v_{\perp}^2}{u_p V_{\rm A}} \right]
  \, = \, - \frac{(u_p + V_{\rm A}) Q_{\rm damp}}{\rho u_p V_{\rm A}}
  \label{eq:action}
\end{equation}
and this equation is solved for $v_{\perp}$, the root mean squared (rms)
transverse velocity amplitude of the MHD fluctuations.
The damping rate $Q_{\rm damp}$ contains a phenomenological description
of the rate of nonlinear cascade for fully-developed MHD turbulence, and
the full expression is given in Equations (18)--(28) of \citet{CvB12}.
Note that $Q_{\rm damp}$ depends nontrivially on the local value
of $v_{\perp}$, so the self-consistent solution for $v_{\perp}(r)$
must be integrated numerically from a specified lower boundary condition.
In the limit of weak damping ($Q_{\rm damp} \rightarrow 0$),
the solution to Equation (\ref{eq:action}) scales as
\begin{equation}
  v_{\perp}^2 \, \propto \, \frac{u_p V_{\rm A}}{(u_p + V_{\rm A})^2}
\end{equation}
which, near the Sun (i.e., where $u_p \ll V_{\rm A}$) behaves as
$v_{\perp} \propto \rho^{-1/4}$ \citep[see also][]{P65,H73}.

Transverse velocity fluctuations can affect the profiles of optically
thin emission lines in several ways:
\begin{enumerate}
\item
Stochastic plasma motions tend to provide a time-averaged Doppler
broadening to a spectral line \citep[see, e.g.,][]{Ma68,Es90}.
This is likely to be related to the observational inference of
unresolved ``microturbulence'' in stellar atmospheres \citep{SE34,Gr73}.
For the solar corona, only the component of the velocity field parallel
to the observer's line of sight (LOS) has a direct influence on the
Doppler broadening.
Thus, we presume that the kinetic energy in the fluctuations is
equipartitioned between the two orthogonal directions transverse to
the background magnetic field.
Only one of those directions influences the spectral line formation,
and the time-averaged LOS component of its velocity amplitude is
specified as $\xi = v_{\perp} / \sqrt{2}$.
The impact of this velocity amplitude on the computed line
profiles is discussed further in Section \ref{sec:emp:Tp}.
\item
MHD fluctuations that propagate through an inhomogeneous background
plasma exert a mean ``wave-pressure'' acceleration on that
plasma \citep{BG68,B71,AC71,HO80}.
In the limiting case of Alfv\'{e}n waves that have equal
kinetic and magnetic energy densities, 
the radial wave-pressure acceleration is specified as
\begin{equation}
  a_{\rm wp} \, = \, -\frac{1}{2\rho} \frac{\partial}{\partial r}
  \left( \rho v_{\perp}^2 \right) \,\, .
\end{equation}
After $v_{\perp}(r)$ has been evaluated numerically for a given
Monte Carlo trial model, $a_{\rm wp}$ is determined by computing the
above radial derivative using centered finite-differencing.
\item
Lastly, at large distances from the Sun, particle collisions and
charge-exchange processes become infrequent enough that the charged
particles respond to MHD flucutations in the magnetic and electric field,
but the neutral atoms do not.
Line profiles produced by neutrals may not be as strongly broadened by
the Doppler motions of the fluctuations as would an equivalent ion line.
However, decoupling between the transverse motions of the neutrals
and the ions may also lead to frictional heating of the neutrals
\citep{Ol94,Al00}.
In existing models, this heating tends to provide an additional amount
of line broadening that is of the {\em same order of magnitude} as the
original Doppler broadening due to ion motions.
Thus, we follow \citet{Cr98} by assuming these two effects
(neutral decoupling and frictional heating) cancel out exactly, and
that the overall broadening of the neutral H~I Ly$\alpha$ line remains
the same as if it were formed by ions.
\end{enumerate}

In order to account for a range of possible results regarding the
MHD velocity fluctuations in the corona, the value of $\xi$ at the
lower boundary of the integration of Equation (\ref{eq:action}) is
varied as one of the Monte Carlo parameters (see Table \ref{tab01}).
At a radial distance $r_0 = 1.4 \, R_{\odot}$, the quantity
$\xi_0$ is selected from a uniform random distribution between
30 and 90 km~s$^{-1}$.
This range was chosen to encompass the uncertainty in the recent
observational literature about the magnitude of nonthermal velocities
measured above the solar limb.
For each Monte Carlo trial model, $v_{\perp}$ is specified at $r_0$,
and Equation (\ref{eq:action}) is integrated up to a maximum radial
height of 5 $R_{\odot}$ using first-order Euler forward differencing.

\begin{figure}[!t]
\epsscale{1.17}
\plotone{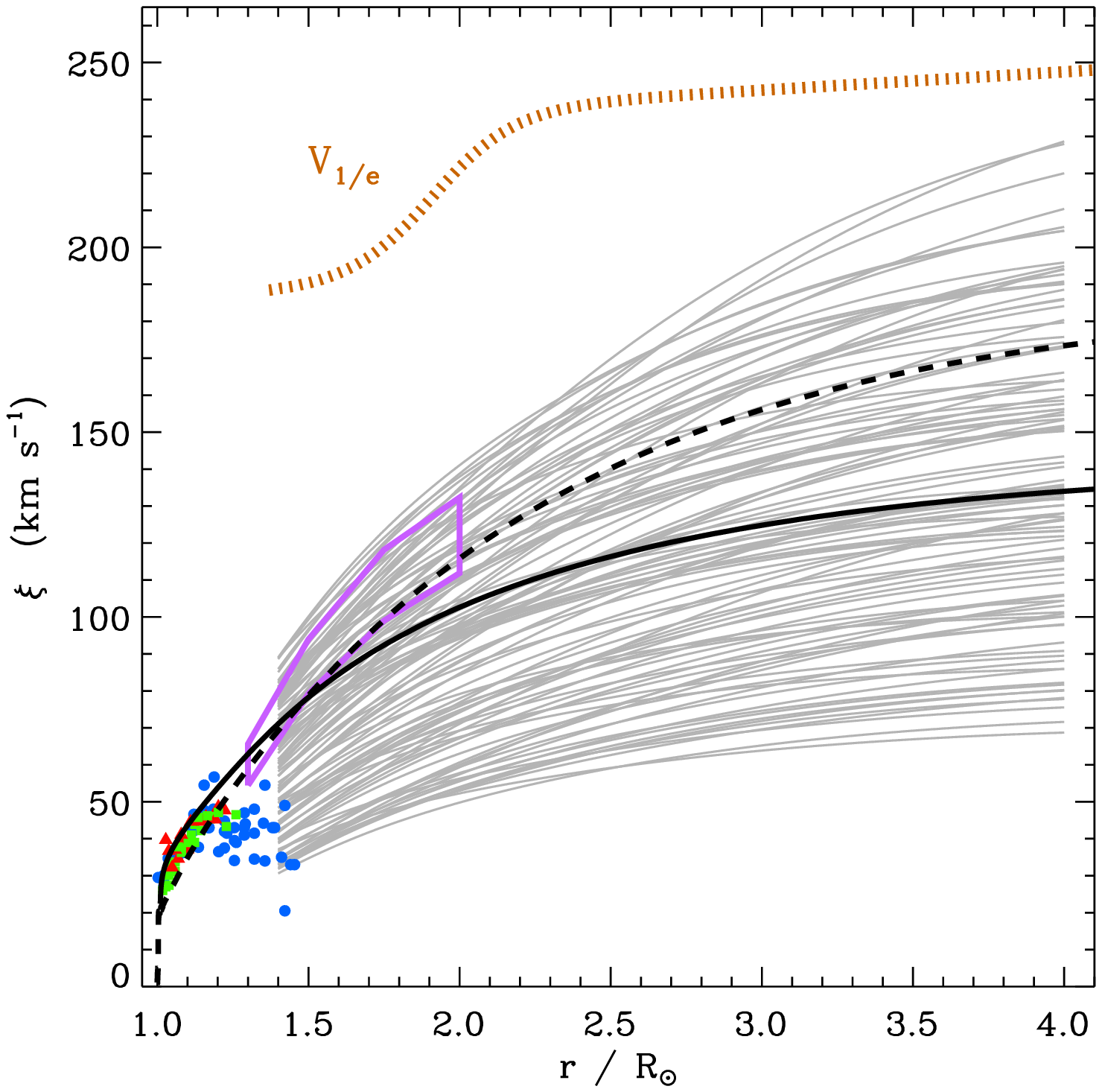}
\caption{Line-of-sight velocity amplitudes $\xi$ shown for a set of
trial Monte Carlo models (thin gray curves), compared with
model results from \citet{CvB05} (dashed thick black curve) and
\citet{CvB07} (solid thick black curve), and with observations from
\citet{Bj98} (green squares) \citet{LC09} (red triangles),
\citet{Hh13} (blue circles), and \citet{Es99} (purple outline).
Also shown is the observed trend in the H~I Ly$\alpha$ line width
from Figure~\ref{fig01} (dotted brown curve), which $\xi$ does
not exceed.
\label{fig04}}
\end{figure}

Figure \ref{fig04} shows a selection of representative $\xi(r)$ curves
and compares them with some observationally inferred values of the
nonthermal velocity \citep{Bj98,Es99,LC09,Hh13}.
For $r < 1.2 \, R_{\odot}$ the measurements largely agree in showing
a monotonically increasing trend for $\xi$ versus height.
Above 1.2~$R_{\odot}$, there is some disagreement.
\citet{Es99} combined multiple line-profile measurements from UVCS
to show how the data were consistent with a lack of damping
(i.e., $v_{\perp} \propto \rho^{-1/4}$).
However, more recent data from {\em Hinode} appear to show substantial
wave dissipation up to the largest heights observable without
occultation \citep{Hh12,BA12,Hh13,Gu17}.
Figure \ref{fig04} illustrates how the full range of possible random
choices for $\xi_0$ (at 1.4~$R_{\odot}$) is meant to allow us to use
the UVCS H~I Ly$\alpha$ data to help resolve this observational tension.

The exact nature of coronal turbulence is not yet understood completely.
Thus, the adopted form of the $Q_{\rm damp}$ term in
Equation (\ref{eq:action}) was only one of many possible descriptions
of the cascade and dissipation process.
One way to assess the verisimilitude of this term is by computing
slopes of the numerically integrated $v_{\perp}(r)$ curves and comparing
them to the results of more sophisticated simulations.
For each model shown in Figure \ref{fig04}, a linear fit was made
to $\log v_{\perp}$ versus $-\log \rho$ over the modeled range of
heights, and the derived slope $m$ is then equivalent to a mean
power-law scaling $v_{\perp} \propto \rho^{-m}$.
Undamped Alfv\'{e}n waves in the corona would have $m = 0.25$.
For the numerical results shown here, the distribution of $m$ values
was roughly normal, with a mean of 0.1352, a standard deviation of 0.0204,
a minimum of 0.074, and a maximum of 0.187.
This range of values is roughly comparable to the output of reduced
MHD simulations.
For comparable heights in the low corona, we extracted power-law
exponents $m$ from these simulations and found values sometimes as
low as 0.12 \citep[from][]{WC15} and sometimes as high as
0.21 \citep[from][]{vB16}.
Neither our $Q_{\rm damp}$ model nor the reduced MHD simulations
ever exhibited values of $m$ as high as the undamped value of 0.25.

\subsection{Anisotropic Hydrogen Temperatures}
\label{sec:emp:Tp}

The shapes and strengths of H~I Ly$\alpha$ profiles depend on the
kinetic velocity distributions of neutral hydrogen atoms in the corona.
In order to accommodate models of coronal heating that use
collisionless physics, we do not assume a purely Maxwellian
velocity distribution.
Instead, a bi-Maxwellian (i.e., two temperature) distribution is
used, with unequal temperatures defined in the directions parallel
and perpendicular to the background magnetic field vector.
As discussed in the previous subsection,
it is also assumed that the neutral hydrogen atoms
exhibit the same velocity distribution as the protons.
The unresolved most-probable speeds in the two orthogonal directions
are specified by
\begin{equation}
  w_{p \parallel}^2 \, = \, \frac{2k_{\rm B}T_{p \parallel}}{m_p}
  \label{eq:wppara}
\end{equation}
\begin{equation}
  w_{p \perp}^2 \, = \, \frac{2k_{\rm B}T_{p \perp}}{m_p} + \xi^2
  \label{eq:wpperp}
\end{equation}
where $k_{\rm B}$ is Boltzmann's constant.
Note that the transverse MHD fluctuations described in
Section \ref{sec:emp:Brxi} are assumed
to vary on spatial and time scales short enough so that it makes sense
to incorporate them directly into the perpendicular most-probable speed.
Sometimes the quantity defined in Equation (\ref{eq:wpperp}) is
expressed in terms of a ``kinetic temperature'' $T_k$, with
$w_{p \perp}^2 = 2 k_{\rm B} T_k / m_p$.

For each trial model in the Monte Carlo ensemble, the hydrogen
most-probable speeds are determined as follows.
First, given that the line profile width is generally known to remain
close to the plane-of-sky value of $w_{p \perp}$, the radial
dependence of the latter is specified as
\begin{equation}
  w_{p \perp} \, = \, \Psi \, V_{1/e} \,\,\, ,
\end{equation}
where $\Psi$ is a dimensionless constant randomly sampled between
0.7 and 1.3.
Equation (\ref{eq:V1efit}) is used to specify the radial variation
of $V_{1/e}$.
Now that both $w_{p \perp}$ and $\xi$ are known,
Equation (\ref{eq:wpperp}) is solved for $T_{p \perp}$.
It is possible that the smallest allowed values for $\Psi$ and the
largest allowed values of $\xi_0$ conspire to produce an unphysically
low value of $T_{p \perp}$.
Thus, at this point, a minimum threshold is applied; i.e., at any
height where $T_{p \perp}$ happens to fall below 0.2~MK, it is set to
be equal to that ``floor'' value.
In the large Monte Carlo ensemble of thousands of random samples,
no more than 7\% of them required this correction.

Once a reasonable model curve for $T_{p \perp}(r)$ exists,
a value for the microscopic temperature anisotropy ratio
$\alpha = T_{p \perp}/T_{p \parallel}$ is sampled between 0.1 and 10,
with the sample taken uniformly in $\log \alpha$.
For simplicity, the quantity $\alpha$ is assumed to be constant as a
function of radial distance.
Such a large dynamic range in $\alpha$ appears to be warranted because
of the wide variety of proton temperature anisotropies measured in
the solar wind \citep[see, e.g.,][]{Ma06,Me07,Ve19}.
The randomly chosen anisotropy ratio is combined with the previously
specified value of $T_{p \perp}$, and thus the full radial dependence of
$T_{p \parallel}$ is obtained straightforwardly.

Together with $n_e(r)$ and $u_p(r)$, the above temperature quantities
are computed on a fine radial grid between 1.4 and 5 $R_{\odot}$.
Values at larger heights are also required in order to specify the
plasma properties along foreground and background parts of the
observational LOS.
The expressions given in Sections \ref{sec:emp:ne}--\ref{sec:emp:Brxi}
can be evaluated safely at heights well above 5~$R_{\odot}$.
However, because of our use of the fitting function for $V_{1/e}$,
the expressions above for $T_{p \parallel}$ and $T_{p \perp}$
should not be extrapolated above 5~$R_{\odot}$.
Thus, for the purposes of LOS integration, it is assumed that both
proton temperatures remain equal to their specified values at
$r = 5 \, R_{\odot}$ at all larger heights.

Although many assumptions have been built into the functional forms of
$T_{p \parallel}(r)$ and $T_{p \perp}(r)$, the eventual set of
validated temperature ``measurements'' is not required to obey these
functions.
Each trial model in the Monte Carlo ensemble may end up agreeing with
the observational data over a limited range of heights.
The proton temperatures that agree with the UVCS data can be sampled
from one subset of Monte Carlo models at $r = 1.5 \, R_{\odot}$,
a different subset at $r = 2.5 \, R_{\odot}$, and a completely
different subset at $r = 4 \, R_{\odot}$.
Thus, completely new radial dependences are free to emerge from
these different subsets.
For example, despite the assumption that the anisotropy ratio
$\alpha$ is constant as a function of radial distance for each model,
the final set of best-fitting values of $\alpha$ can in fact exhibit
some radial variation (see, e.g., Figure \ref{fig10}(a) below).

\subsection{Electron Temperature}
\label{sec:emp:Te}

The resonantly scattered H~I Ly$\alpha$ line depends mainly on
the properties of neutral hydrogen atoms, but it was seen in
Section \ref{sec:emp:ne} that the number density $n_{\rm HI}$ also
depends on the electron temperature $T_e$.
At large heights in polar coronal holes, the plasma is collisionless
enough that the protons and electrons are not guaranteed to be in
thermal equilibrium.
Thus, we require some additional observational or model-based
constraints on $T_e$.

Unfortunately, reliable electron temperature data do not appear to
exist for off-limb heights above a few tenths of a solar radius
in coronal holes.
There are some extreme-ultraviolet collisional line-pairs that are
sensitive to $T_e$ \citep[e.g.,][]{Fd99,Ds01,Wm06,La08},
but coronagraphic occultation would be needed to extend those
observations to the heights sampled by UVCS.
X-ray filter ratios have been applied to off-limb solar data
\citep[e.g.,][]{Fo97,AA01}, but count rates are low in coronal holes
and the interpretation of LOS-integrated $T_e$ values is difficult.
``Frozen-in'' ion charge states measured at 1~AU have been used to
infer coronal electron temperatures
\citep{Ow83,YKo97,Ae98,EE00,Ln14}, but the results depend
strongly on assumptions about the shape of the velocity distribution
function and differential flow between the ions.
The shape of the Thomson-scattered visible-light continuum can be
used to measure $T_e$ \citep{Cram76,Rg00,Rg11},
but existing measurements from total eclipses still have large
uncertainties for dark coronal-hole regions.
Lastly, the Thomson-scattered component of the H~I Ly$\alpha$
line itself is sensitive to $T_e$, but that component is exceedingly
dim as well \cite[see, e.g.,][]{Hu65,Wi82,Fi98}.

\begin{figure}[!t]
\epsscale{1.20}
\plotone{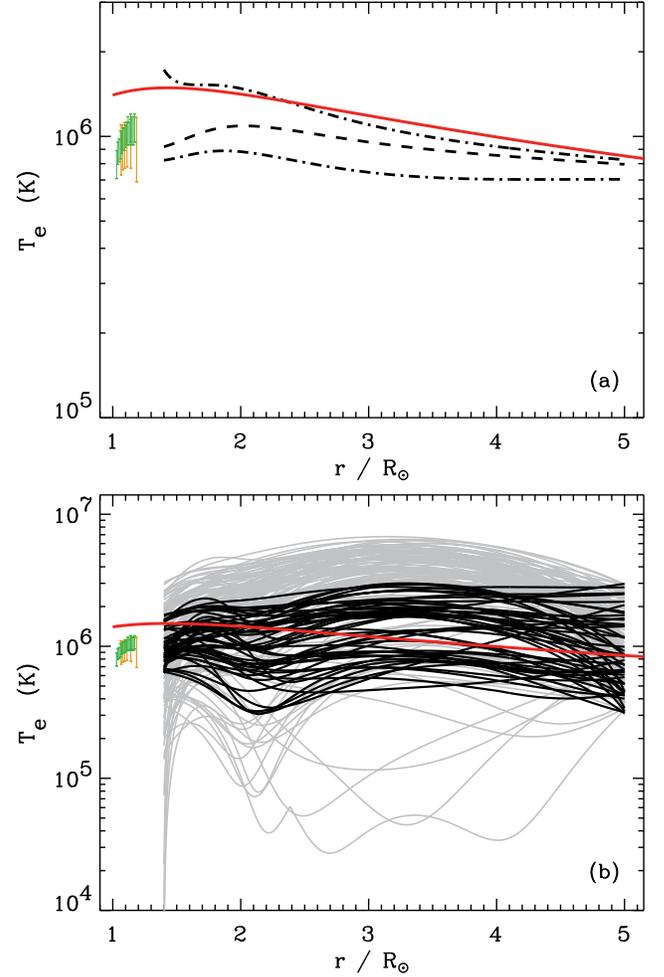}
\caption{Electron temperatures derived from momentum conservation
equations:
(a) Hydrostatic solutions of Equation (\ref{eq:static}) for $n_{e,0}$
(dashed black curve) and two extreme choices of $\Delta_{\rm lo}$
and $\Delta_{\rm hi}$ (dot-dashed black curves; see text).
(b) Solar-wind solutions of Equation (\ref{eq:momentum}), both with
(black curves) and without (gray curves) the initial cut for marginal
consistency with in~situ charge-state data.
Both panels show observationally inferred temperatures from
\citet{YKo97} (red solid curve), \citet{La08} (green error bars),
and \citet{Wm06} (gold error bars).
\vspace*{0.30in}
\label{fig05}}
\end{figure}

Another traditional way of determining $T_e$ is to use the radial
momentum conservation equation.
In hydrostatic equilibrium, one can assume the absence of a radial flow,
no wave pressure, and an isotropic one-fluid temperature that varies
with radius much more slowly than $n_e$ \citep[e.g.,][]{A41,FG95}.
Under those assumptions, the ``scale-height temperature'' is given by
\begin{equation}
  T_e \, \approx \, \frac{GM_{\odot} m_p}{2 k_{\rm B} r^2}
  \left| \frac{n_e}{\partial n_e / \partial r} \right|
  \label{eq:static}
\end{equation}
where $G$ is Newton's gravitational constant and $M_{\odot}$ is
the solar mass.
Figure \ref{fig05}(a) shows solutions to this equation both for the
floor density $n_{e,0}$ and for two extreme choices of the other
density parameters that
maximize ($\Delta_{\rm lo} = 1$, $\Delta_{\rm hi} = 4$)
and minimize ($\Delta_{\rm lo} = 4$, $\Delta_{\rm hi} = 1$)
the hydrostatic temperature.

A more general version of the momentum equation includes the presence
of a mass-conserving solar wind \citep{MJ77,LS16,LK20}, wave-pressure
acceleration, and anisotropic proton temperatures.
Assuming all particle species flow together at a common bulk speed
$u_p$, the single-fluid momentum equation for bi-Maxwellian protons
and isotropic Maxwellian electrons is
\begin{displaymath}
  u_p \frac{\partial u_p}{\partial r} \, = \,
  - \frac{k_{\rm B}}{n_p m_p} \frac{\partial}{\partial r}
  \left[ n_p \left( T_{p \parallel} + T_e \right) \right]
\end{displaymath}
\begin{equation}
  + \, \frac{k_{\rm B}}{A m_p} \frac{\partial A}{\partial r}
  \left( T_{p \perp} - T_{p \parallel} \right)
  - \frac{GM_{\odot}}{r^2} + a_{\rm wp}  \,\, .
  \label{eq:momentum}
\end{equation}
Note that this reduces to Equation (\ref{eq:static}) in the limit
of $u_p = 0$, $a_{\rm wp} = 0$, $T_{p \parallel}=T_{p \perp}=T_e$,
and the assumption that the temperature can be removed from the
pressure-gradient derivative.
In the present case, solving Equation (\ref{eq:momentum}) for $T_e$
requires numerical integration from a specified boundary condition.
A value for $T_e$ was specified at an upper boundary of 5~$R_{\odot}$,
and Equation (\ref{eq:momentum}) was integrated down to
1.4~$R_{\odot}$ using first-order Euler differencing.
The boundary temperature $T_{\rm top}$ is the final Monte Carlo
parameter listed in Table~\ref{tab01}, and it was sampled from a
uniform distribution between 0.3 and 3 MK.
When integrating down from different values of $T_{\rm top}$,
\citet{LS16} found that the $T_e(r)$ curves converge toward a common
solution and the choice of upper boundary condition matters less
at lower heights.

Figure \ref{fig05}(b) shows a selection of 170 trial solutions to
Equation (\ref{eq:momentum}).
Each solution involves randomly sampling from all 8 parameters in
Table~\ref{tab01}, and thus it gives rise to a huge range of possible
electron temperatures.
Over all possible combinations of parameters, the maximum value of
$T_e$ is about 8~MK, and the minimum values may become negative.
Thus, an initial plausibility criterion was imposed to maintain
marginal consistency with the existing range of data for $T_e$
discussed above and shown in Figure \ref{fig05}.
Only models with:
(1) a minimum $T_e$ greater than 0.3 MK,
(2) a maximum $T_e$ less than 3.0 MK,
(3) a value of $T_e$ at the base ($r = 1.4 \, R_{\odot}$) between
0.6 and 1.8 MK,
were found to be plausible enough to move forward in the Monte Carlo
ensemble.
For the 170 trial cases shown in Figure \ref{fig05}(b), 49 of them
satisfied the above criteria and are shown as black curves.

\section{Simulating Lyman Alpha Emission}
\label{sec:lya}

Each random trial in the Monte Carlo ensemble requires calculation
of the H~I Ly$\alpha$ emission line profile over a range of
observation heights.
For the heights sampled by UVCS, this emission line is formed
primarily by resonance scattering of photons by neutral hydrogen
atoms in the extended corona \citep{Gb71,BC74,Wi82,NM99}.
Here, it is safe to ignore
H~I Ly$\alpha$ emission from collisional processes and Thomson
scattering, and also to model the resonance scattering in the limit
of the \citet{H62} Case~I approximation \citep[see also][]{Cr98}.
The general expression for the specific intensity along the
observational LOS direction $\hat{\bf n}$ is
\begin{displaymath}
  I_{\nu} (\hat{\bf n}) \, = \,
  \frac{h \nu_0}{4\pi} B_{12} \int dx \,\, n_{\rm HI} \,\, \times
\end{displaymath}
\begin{equation}
  \int d\nu' \oint \frac{d\Omega'}{4\pi} \,
  {\cal R} (\nu', \hat{\bf n}', \nu, \hat{\bf n}) \,
  I_{\odot \nu'}(\hat{\bf n}') \,\, ,
  \label{eq:InuLOS}
\end{equation}
where $\nu_0$ is the rest-frame frequency of the line, $B_{12}$ is
its Einstein absorption rate, and $x$ is the coordinate parallel to
the LOS.
The Case~I redistribution function ${\cal R}$ depends not only on
the observed frequency $\nu$, but also on the incident frequency
$\nu'$ from the solar disk, the incident direction vector $\hat{\bf n}'$,
and the properties of the neutral hydrogen velocity distribution,
which we parameterize using $u_p$, $w_{p \parallel}$, and $w_{p \perp}$.
Generally, for larger values of $u_p$, there are fewer photons
scattered into the observer's LOS; this phenomenon is called
{\em Doppler dimming.}
The emission at each point along the LOS depends on an integration
over all incident frequencies and directions from the solar disk.
The latter is a two-dimensional integral over solid angle, which
we express in spherical coordinates as
$d\Omega' = \sin\theta' \, d\theta' \, d\phi'$.
Additional details about Equation (\ref{eq:InuLOS}) are given by,
e.g., \citet{Cr98} and \citet{Cr99}.

Because the relevant H~I Ly$\alpha$ photons are emitted initally
from the upper chromosphere then scattered into our LOS, the
observed profiles depend on the intensity $I_{\odot \nu'}$
coming from the solar disk.
The shape of this incident spectral energy distribution was taken
from the atlas obtained by the Solar Ultraviolet Measurements of
Emitted Radiation (SUMER) instrument on {\em SOHO} \citep{Cu01}.
Note that some earlier modeling efforts \citep[e.g.,][]{Cr99}
used only a narrow range of $\pm 1$~{\AA} around the rest-frame
wavelength and assumed $I_{\odot \nu'} = 0$ outside that range.
Here, we found that some of the Monte Carlo models involve quite
large Doppler shifts, so we kept a larger range of
$\pm 13$~{\AA} around the rest-frame wavelength for safety.

\begin{figure}[!t]
\epsscale{1.17}
\plotone{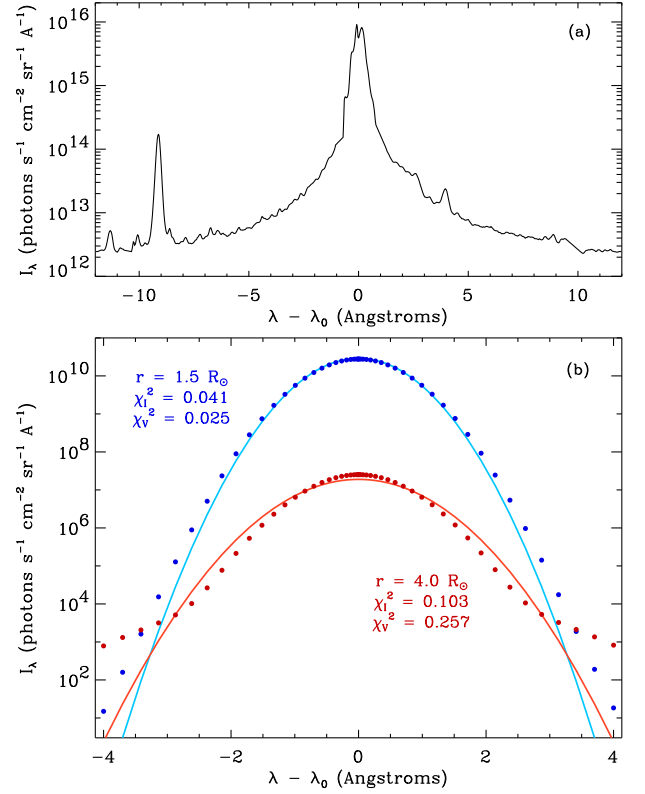}
\caption{(a) Solar-disk spectral energy distribution in the
vicinity of H~I Ly$\alpha$, with $\lambda_0 = 1215.67$~{\AA}.
(b) Comparison of simulated profiles (points) and reconstructed
Gaussian estimates from observed values of $I_{\rm tot}$ and
$V_{1/e}$ (solid curves) at heights of $r = 1.5 \, R_{\odot}$
(blue) and $4 \, R_{\odot}$ (red).
For the $r = 1.5 \, R_{\odot}$ model, parameters were:
$\Delta_{\rm lo} = 1.233$,
$\Delta_{\rm hi} = 3.886$,
$u_{\infty} = 463.6$ km~s$^{-1}$,
$f_{\rm max} = 7.701$,
$\xi_0 = 64.19$ km~s$^{-1}$,
$\Psi = 0.9215$,
$\alpha = 0.9207$,
$T_{\rm top} = 0.8710$ MK.
For the $r = 4 \, R_{\odot}$ model, parameters were:
$\Delta_{\rm lo} = 1.035$,
$\Delta_{\rm hi} = 2.741$,
$u_{\infty} = 712.1$ km~s$^{-1}$,
$f_{\rm max} = 9.790$,
$\xi_0 = 42.37$ km~s$^{-1}$,
$\Psi = 1.218$,
$\alpha = 0.5116$,
$T_{\rm top} = 0.8166$ MK.
\label{fig06}}
\end{figure}

The Sun's H~I Ly$\alpha$ intensity varies substantially over the
solar cycle, so the intensities in the SUMER atlas were scaled to
agree with the times of the relevant UVCS data.
Below, values of the total integrated intensity $I_{15}$ are specified
in units of $10^{15}$ photons s$^{-1}$ cm$^{-2}$ sr$^{-1}$.
\citet{Cr99} used UVCS on-disk measurements of \citet{Ra97} from
1996 December 4, which found $I_{15} = 5.24$.
\citet{Nk08} took an average over the 1996--1997 solar minimum
and found $I_{15} \approx 4.5$.
The \citet{Wo00} spectral irradiance database gave mean values of
$I_{15} = 5.167$ for the 1996 December 28 to 1997 January 5 (CR 1917)
time period, and
$I_{15} = 5.260$ for the 1997 April 14 to 1997 April 20 (CR 1921)
time period.
The average of those two values, $I_{15} = 5.214$, is used in the
H~I Ly$\alpha$ simulations for this paper's Monte Carlo ensemble,
and Figure \ref{fig06}(a) shows the SUMER spectrum normalized to
this adopted total intensity.

In order to compare with the UVCS data shown in Figure \ref{fig01},
the H~I Ly$\alpha$ emission lines were simulated using the CUES code
\citep[described by][]{Cr99,Cr08} for a grid of 28 lines of sight
with impact parameters $r$ between 1.4 and 4.1 $R_{\odot}$ in steps
of 0.1 $R_{\odot}$.
Each profile was simulated with 51 wavelength points spread between
$-4$~{\AA} and $+4$~{\AA} of the rest-frame wavelength of
$\lambda_0 = 1215.67$~{\AA}.
To have higher spectral resolution near line-center, a stretched
wavelength grid was specified.
First, a uniform grid was created in a dimensionless quantity $q$,
with $-1 \leq q \leq +1$.
Then, the transformation to wavelength was performed using
\begin{equation}
  \lambda \, = \, \lambda_0 + q |q|^{0.9} (\mbox{4~{\AA}}) \,\, ,
\end{equation}
so that, for a grid of 51 points, the grid spacing at line-center is
only 0.0088~{\AA}, but at the edges it grows to 0.2985~{\AA}.
The integration over the LOS-direction $x$ was discretized with
a fixed step-size $\Delta x = 0.15 \, R_{\odot}$, and the integration
limits in $x$ were set to be $\pm 8 r$.
In other words, as the observation height increases, the absolute
range of $x$ values included in the integral also increases.

When analyzing the results from each Monte Carlo trial model, the
simulated lines were fit by Gaussian functions.
Nonlinear least-squares fitting was performed with the GAUSSFIT
routine of the Interactive Data Language (IDL), which uses a
gradient-expansion algorithm \citep[see, e.g.,][]{Mq63}.
As described in more detail in Section \ref{sec:results}, sometimes
the simulated profiles exhibited central reversals and were not
well-fit by a single Gaussian.
These cases end up being ruled out (prior to fitting) because this
behavior is not seen in the UVCS coronal-hole data.

Each simulated LOS is assumed to pass directly over the polar axis of
an axisymmetric coronal hole.
Because of the superradial expansion described in
Section \ref{sec:emp:up}, the solar wind velocity vector does not
always point radially away from the Sun.
At points along the foreground and background of any given LOS, there
tends to be an extra deflection away from the polar axis by about
5$^{\circ}$ to 15$^{\circ}$ \citep[see, e.g., Figure 2 of][]{Cr08}.
The shapes of magnetic field lines inside the coronal hole are assumed
to self-similarly follow polar angles $\theta$ that remain proportional
to the outer boundary of the coronal hole as a function of radial
distance.
This is the same assumption made by \citet{Cr99}, but in this
case the expansion factor $f(r)$ is given by Equation (\ref{eq:fsuper}).

Two examples of simulated H~I Ly$\alpha$ emission
lines are shown in Figure \ref{fig06}(b).
Modeled profiles are shown with discrete points, and idealized
Gaussian reconstructions of the {\em observed} data at two heights
are shown with solid curves.
Despite these profiles not being formal fits to the simulations,
we nevertheless compute and show $\chi^2$-like goodness-of-fit
parameters (see below for definitions) to illustrate how specific
values of these quantities correspond to various levels of agreement.
The simulated profile at $r = 4 \, R_{\odot}$ exhibits a clear
power-law departure from a Gaussian shape in the line wings.
This effect arises because line broadening due to bulk solar wind
outflow does not behave in the same way as more random sources of
``nonthermal'' broadening like waves or microturbulence
(see, e.g., \citeauthor{Ko06tail} \citeyear{Ko06tail}, as well as
\citeauthor{GC20} \citeyear{GC20}).
It is worth repeating that the line-synthesis procedure assumes
the bulk outflow speed of the neutral hydrogen atoms is equal to the
bulk outflow speed $u_p(r)$ of the protons.

\section{Results: Electron and Proton Plasma Properties}
\label{sec:results}

\begin{figure}[!t]
\epsscale{1.17}
\plotone{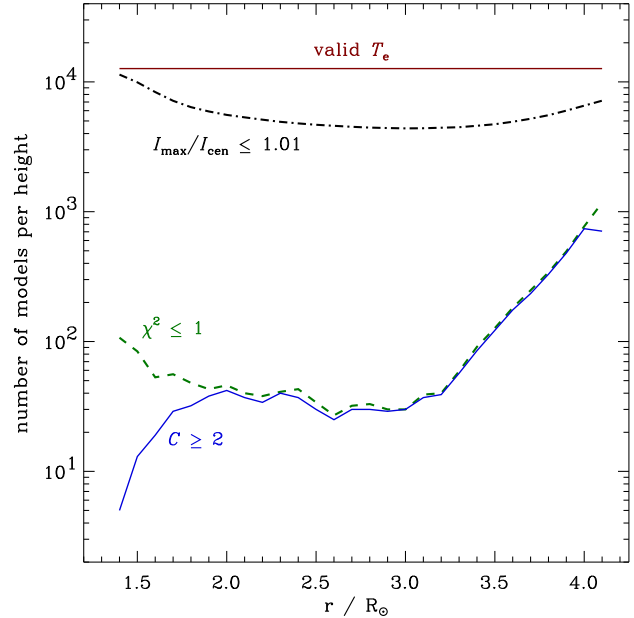}
\caption{Numbers of simulated H~I Ly$\alpha$ profiles in the Monte
Carlo ensemble that passed each of the four successive tests
described in the text:
consistency with observed $T_e$ values (brown solid curve),
lack of a central reversal in the simulated profile (black
dot-dashed curve), agreement with observed intensities and line
widths within $1\sigma$ (green dashed curve), and radial coherency
(blue solid curve).
\label{fig07}}
\end{figure}

Because there are 8 independently varied Monte Carlo variables
(see Table \ref{tab01}), there is an 8-dimensional volume of
parameter space that must be searched for unique combinations of
plasma properties that match the UVCS data.
However, it is somewhat computationally intensive to run the CUES
line-synthesis code.
Thus, a tradeoff had to be found between a finely exhaustive search
of the 8-dimensional parameter space and calculations that could be
completed in a reasonable time.
We ended up creating $10^5$ random trial sets of the 8 Monte Carlo
parameters, and we put them through a four-step winnowing process
to determine whether a given model can be claimed to ``agree''
with the observations:
\begin{enumerate}
\item
As described in Section \ref{sec:emp:Te}, any trial model with
values of $T_e$ that fell outside a range of marginal consistency
with existing near-limb and charge-state data was rejected.
Out of the $10^5$ initial sets of trial parameters, only
12,657 sets satisfied this consistency criterion (i.e., 12.657\%).
\item
All remaining models were run through CUES to compute 354,396
individual H~I Ly$\alpha$ profiles (i.e., 28 observation heights for
each set of parameters that survived the previous step).
In the UVCS data, off-limb emission lines in coronal holes were
always clearly single-peaked; i.e., there was a global intensity
maximum at line-center, and the coronal profiles tended to be
well-fit by a single Gaussian function.
Thus, all nonmonotonic profiles were rejected from further
consideration.
The ratio $I_{\rm max}/I_{\rm cen}$ was used to judge nonmonotonicity,
where $I_{\rm max}$ is the maximum value of the specific intensity
across the entire line profile, and $I_{\rm cen}$ is the value of
the specific intensity at its centroid (first moment) wavelength.
Usually the centroid wavelength falls within 0.1~{\AA} of the
rest wavelength $\lambda_0$, but not always.
Synthesized profiles with a value of $I_{\rm max}/I_{\rm cen}$
exceeding 1.01 were judged to exhibit a non-Gaussian central
reversal, so they were excluded from further consideration.
Of the 354,396 synthesized profiles, 159,849 survived this cut
(i.e., only 5.709\% of the initial set of all trials).
\item
Next, Gaussian fits were made to the remaining profiles.
The widths $V_{1/e}$ and total intensities $I_{\rm tot}$
were compared with the UVCS data at the same observation height
$r$ (see Figure \ref{fig01}).
Separate goodness-of-fit quantities were computed for the total
intensity
\begin{equation}
  \chi_{\rm I}^2 \, = \, \left(
  \frac{I_{\rm tot,fit} - I_{\rm tot,obs}}{\delta I_{\rm tot,obs}}
  \right)^2
\end{equation}
and for the line width
\begin{equation}
  \chi_{\rm V}^2 \, = \, \left(
  \frac{V_{1/e,{\rm fit}} - V_{1/e,{\rm obs}}}
  {\delta V_{1/e,{\rm obs}}} \right)^2
\end{equation}
where Equations (\ref{eq:Itotfit})--(\ref{eq:V1efit}) were used
to specify $I_{\rm tot,obs}$ and $V_{1/e,{\rm obs}}$.
Linear interpolation of the curves in Figure \ref{fig01}(c) was
used to specify the uncertainties
$\delta I_{\rm tot,obs}$ and $\delta V_{1/e,{\rm obs}}$.
A useful criterion for a ``good fit to the data'' was found
to be $\max ( \chi_{\rm I}^2 , \chi_{\rm V}^2 ) \leq 1$.
In other words, if the worst-performing of the two criteria 
still showed a variation {\em inside} the $\pm$1$\sigma$ error bars,
it was judged to be an acceptable trial model for this particular
observation height.
Of the 159,849 profiles that made it this far, only 4,350 satisfied
this criterion (i.e., 0.155\% of the initial set of all trials).
\item
Lastly, it was observed that a few of the trial models exhibited
good agreement with the data at one particular height, but the radial
gradients of, say, $I_{\rm tot,fit}$ and $I_{\rm tot,obs}$ were
very different from one another.
This one point of agreement was just a coincidental crossing.
These solutions were found to be unrealistic, so a final coherency
criterion was applied.
For each profile with $\max ( \chi_{\rm I}^2 , \chi_{\rm V}^2 ) \leq 1$,
a coherency parameter $C$ was computed, where $C$ is the number of
directly neighboring profiles---at both lower and higher values of
$r$---that also have $\max ( \chi_{\rm I}^2 , \chi_{\rm V}^2 ) \leq 1$.
A profile that coincidentally agrees with the data at one height
but not any neighboring heights would have $C=1$.
We found that keeping all profiles with $C \geq 2$ successfully
eliminated the few pathological cases.
The final number of remaining profiles was 3,507
(i.e., 0.125\% of the initial set of all trials).
\end{enumerate}
\noindent
Figure \ref{fig07} shows how many profiles were retained at each
of the 28 observation heights and at each of the above steps of the
winnowing process.

The Monte Carlo model parameters for the validated subset of trials
allowed the construction of radially dependent {\em posterior probability
distributions} for plasma parameters such as $n_e$, $u_p$, $B_r$,
$V_{\rm A}$, $v_{\perp}$, $T_{p \parallel}$, $T_{p \perp}$, and $T_e$.
The remainder of this section presents these distributions as a
function of observation height $r$.
It is evident from these distributions that the minimum and maximum
values for the original parameters---given in Table \ref{tab01}---were
sufficiently broad to allow a full exploration of the 8-dimensional
parameter space.
None of the posterior distributions were found to be significantly
edged up against the boundaries of parameter space.
Thus, we can be somewhat confident that the validated ``measurements''
were not limited by the initial choices of model parameters.

\begin{figure}[!t]
\epsscale{1.19}
\plotone{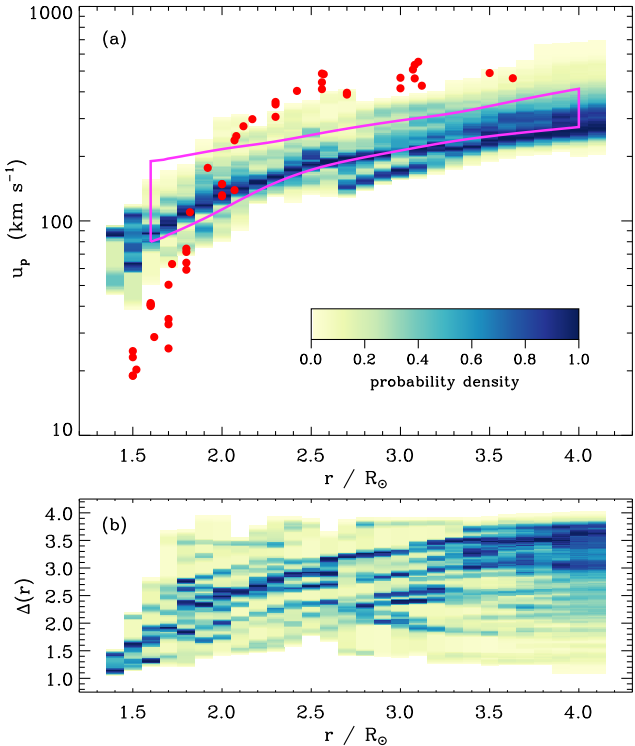}
\caption{Monte Carlo posterior distributions for: (a) radial outflow
speed $u_p$ and (b) the electron density multiplier $\Delta$.
Normalized probability densities are shown with a yellow-green-blue
(viridis-style) color map.
Also shown are proton outflow speeds from
\citeauthor{Cr99} (\citeyear{Cr99}, purple outline) and
O$^{+5}$ outflow speeds from
\citeauthor{Cr08} (\citeyear{Cr08}, red symbols).
\vspace*{0.20in}
\label{fig08}}
\end{figure}

Figure \ref{fig08} shows the distributions of $u_p(r)$ and $\Delta(r)$
for the 3,507 fully validated cases.
At each radial distance $r$, the probabilities were independently
normalized to a common maximum value of 1 (i.e., the darkest blue
color) so the regions of highest probability are always visible by eye.
Note that the distributions appear smoother at the largest heights,
where there were several hundred validated models at each discrete
observation height.
As indicated in the lower left of Figure \ref{fig07}, the smallest
observation height of 1.4~$R_{\odot}$ had only 5 validated models.
Thus, the distributions there are more random and irregular.
The main reason for there being more validated models at larger
heights is that the observational uncertainties at those
heights were larger (see, e.g., Figure \ref{fig01}(c)).

The collection of validated solar-wind velocities shown in
Figure \ref{fig08}(a) agrees well with the proton bulk outflow
velocities from \citet{Cr99}.
This is not surprising, since the analysis of \citet{Cr99} was based
on the same set of UVCS/{\em{SOHO}} H~I Ly$\alpha$ data as was
used in this paper, but the analysis procedure was quite different.
These values are also quite similar to other published results
\citep[see, e.g.,][]{Zn02,Bm17,Sp17,Do18} for comparable ranges
of heights above polar coronal holes.
Also shown in Figure~\ref{fig08}(a) are observationally inferred
O$^{+5}$ ion outflow velocities from \citet{Cr08}.
The heavy ions appear to be flowing substantially slower than the
protons at low heights and faster than the protons at larger heights.

Figure \ref{fig08}(b) shows that $\Delta(r)$ appears to be
increasing from 1.4 to 4~$R_{\odot}$, which is somewhat similar
to the ZEPHYR model result shown in Figure \ref{fig02}(b).
The validated distributions for other ingredients to the mass-flux
conservation equations from Section \ref{sec:emp:up} were
not as clear-cut.
For example, in the full set of 3,507 validated models,
$f_{\rm max}$ appeared to be distributed uniformly between
the specified minimum and maximum (i.e., there was no clear peak in
the distribution of values).
For the $u_{\infty}$ model parameter, there seemed to be a mild
preference for intermediate values; the validated distribution
had a median value of 463 km~s$^{-1}$ and a
standard deviation of 93 km~s$^{-1}$.
This is lower than expected for the fast solar wind associated
with polar coronal holes, but it is also true that the UVCS data
put no real constraints on the existence of continued acceleration
above 4.1~$R_{\odot}$.

\begin{figure}[!t]
\epsscale{1.17}
\plotone{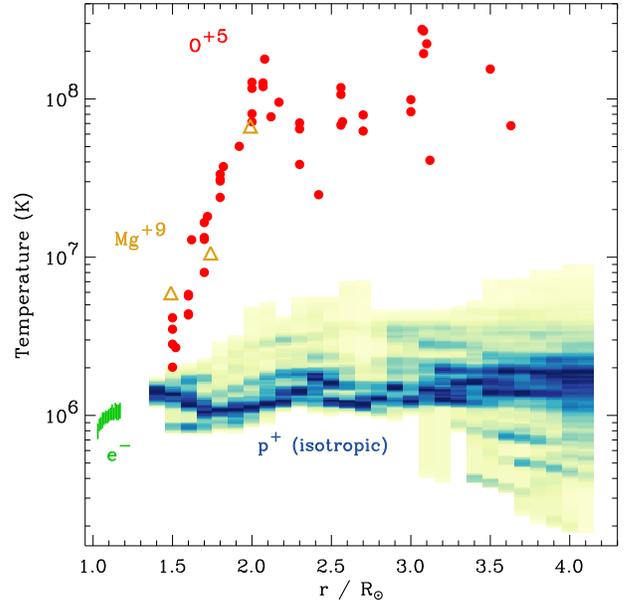}
\caption{Plasma temperatures inferred from UVCS and other data.
Monte Carlo posterior distributions for $T_p$ are shown with the
same color map as Figure \ref{fig08}.
Also shown are perpendicular ion temperatures for O$^{+5}$ 
(red circles) and Mg$^{+9}$ (orange triangles), as well as
$T_e$ data from \citet{La08} (green error bars).
\label{fig09}}
\end{figure}

\begin{figure}[!t]
\epsscale{1.17}
\plotone{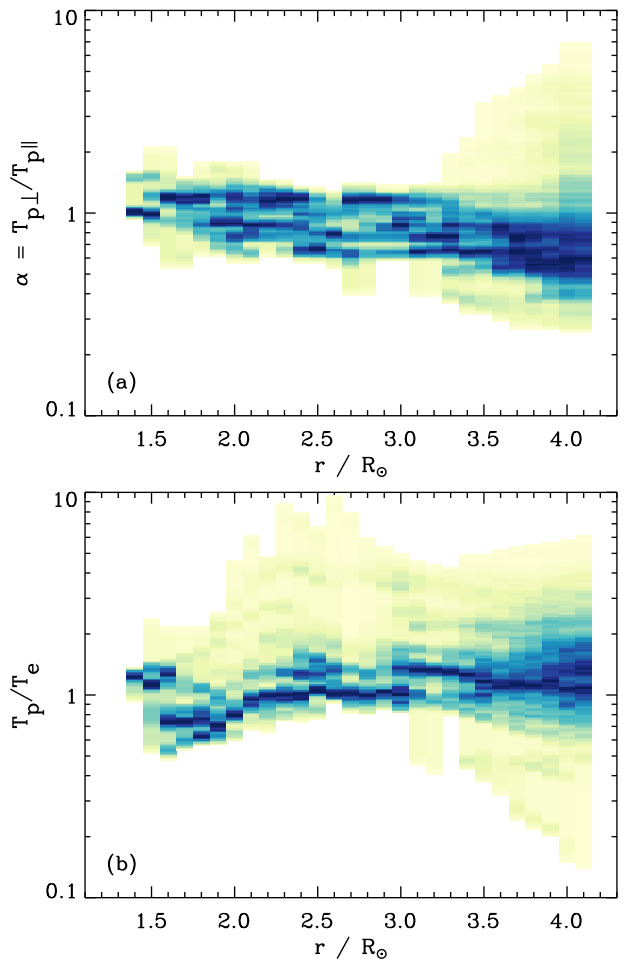}
\caption{Monte Carlo posterior distributions for:
(a) the proton temperature anisotropy ratio
$T_{p \perp}/T_{p \parallel}$, and
(b) the ratio of isotropic proton and electron temperatures
$T_p / T_e$, using the same color map as Figure \ref{fig08}.
\label{fig10}}
\end{figure}

\begin{figure}[!t]
\epsscale{1.17}
\plotone{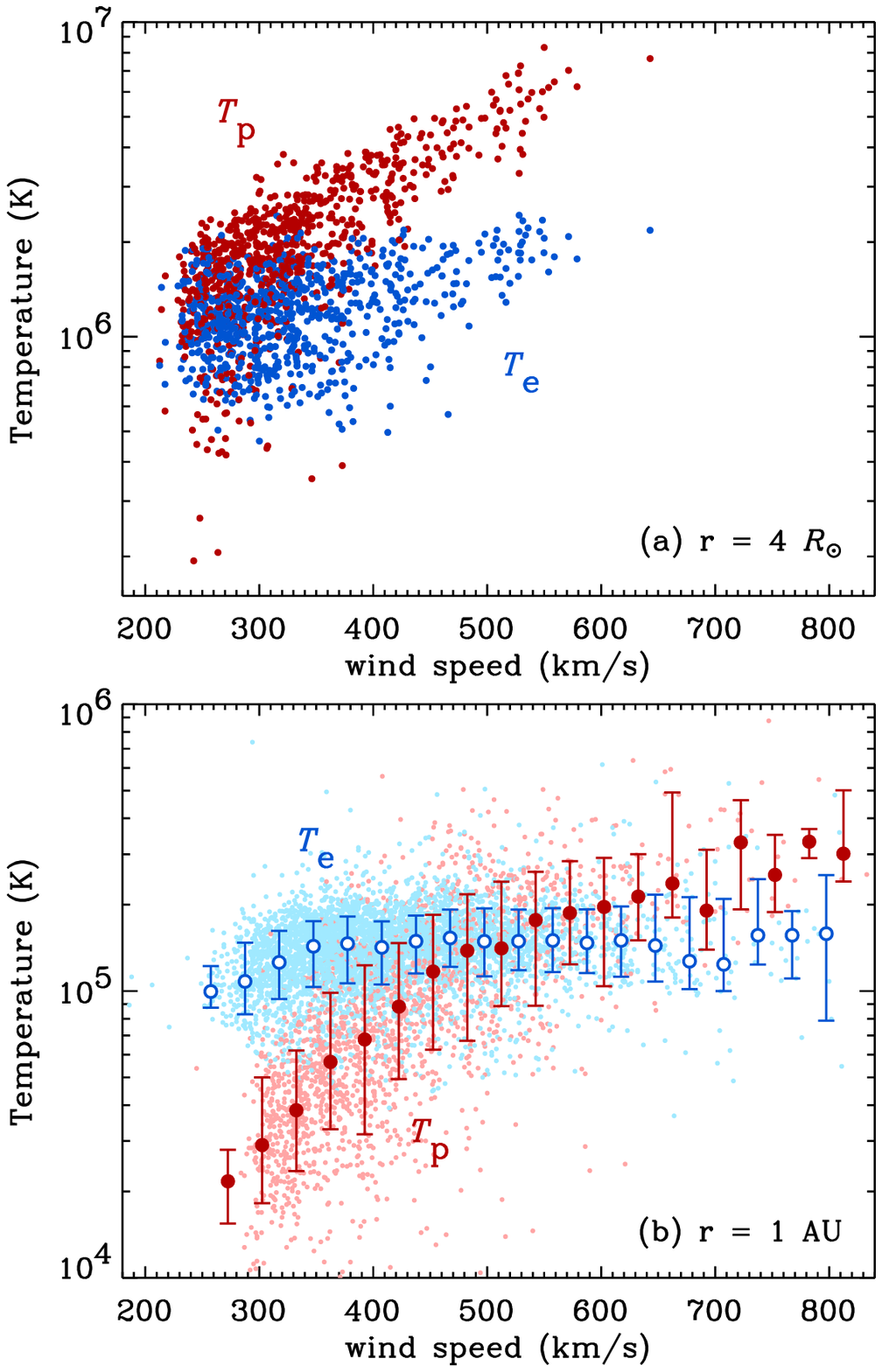}
\caption{(a) Full set of validated Monte Carlo results for $T_p$
(red points) and $T_e$ (blue points) as a function of $u_p$ at
$r = 4 \, R_{\odot}$.
(b) Hourly averaged proton (red) and electron (blue) temperatures
measured at 1~AU by {\em ISEE-3} from January 1980 to October 1982.
Small points indicate individual measurements, and large symbols
show median and $\pm 1 \sigma$ values in 30 km~s$^{-1}$ bins of
wind speed.  See Figure~3 of \citet{Cr17} for additional details.
\label{fig11}}
\end{figure}

Results for validated coronal temperatures are shown in
Figures \ref{fig09}, \ref{fig10}, and \ref{fig11}.
Specifically, the posterior distributions for the input model parameters
(i.e., $\alpha$, $\Psi$, and $T_{\rm top}$) showed no clear trends
because, as expected, each trial model only agreed with the UVCS data
for a limited range of heights.
The resulting distributions for $T_{p \parallel}(r)$,
$ T_{p \perp}(r)$, and $T_e(r)$ tell a more consistent story.
Figure~\ref{fig09} shows radially dependent distributions for
the isotropic proton temperature, defined as
\begin{equation}
  T_p \, = \, \frac{T_{p \parallel} + 2 T_{p \perp}}{3} \,\, ,
\end{equation}
and it seems to be well-constrained to remain between about
1 and 2 MK over the observed heights.
It is important to note that many earlier UVCS studies of proton
``kinetic temperatures'' in coronal holes reported higher values
of order 3--4 MK.
As shown in Equation (\ref{eq:wpperp}), estimation of the temperature
from the line-width alone does not disambiguate between Doppler
broadening due to random thermal motions and that due to unresolved
waves or turbulence.
An earlier attempt to subtract a specific model of Alfv\'{e}nic
turbulence from measured H~I Ly$\alpha$ kinetic temperatures
was presented in Figure~6 of \citet{Cr09}, and the result was a
range of values between 1 and 2 MK, similar to what is shown here.

For comparison to the proton temperatures, Figure~\ref{fig09} also
shows perpendicular ion temperatures derived from the
O~VI 1032,~1037~{\AA} doublet \citep{Cr08}
and from the Mg~X 625~{\AA} line \citep{Ko99}.
Corresponding parallel temperatures for these ions either have
not yet been computed (for Mg~X) or they have large uncertainties
that are difficult to show as error bars (for O~VI).
To convert the kinetic temperatures to true thermal temperatures
in this figure, a simple power-law model for $\xi(r)$ has been
subtracted from each O~VI and Mg~X data point;
see Equation (\ref{eq:xifit}) below.

Figure~\ref{fig10} shows validated distributions for the
anisotropy ratio $\alpha = T_{p \perp}/T_{p \parallel}$ and the
ratio of isotropic proton to electron temperatures $T_p / T_e$.
Although some mild trends in these quantities can be seen as a function
of observation height, they are both also marginally consistent
with being {\em constant} between 1.4 and 4.1~$R_{\odot}$.
Taking the parameters for the full set of 3,507 validated profiles
(covering all heights),
the median values of $T_{p \parallel}$, $T_{p \perp}$,
and $T_e$ are 1.755 MK, 1.918 MK, and 1.236 MK, respectively.
Interestingly, the median anisotropy ratio that one obtains from the
above values is $\alpha = 1.093$, but the median of the original set
of 3,507 values of $\alpha$ itself is 0.9424.
This discrepancy appears to be the result of there being many more
validated models at the largest heights (where $\alpha$ tends to
be less than 1) than at the lowest heights.
There is a similar discrepancy for the $T_p/T_e$ ratio, but it is
much smaller; i.e., using the above median temperatures gives
$T_p/T_e = 1.508$, but the median of the distribution of
$T_p/T_e$ values is 1.501.

The results shown above have implications on theories of proton
and electron heating; these are discussed in more detail in
the Appendix.
For now, we only discuss one additional trend in the
proton--electron disequilibrium (i.e., $T_p \neq T_e$) that
appears to resemble the behavior of the solar wind at 1~AU.
Figure \ref{fig11}(a) examines the 740 validated model parameters
for $T_p$ and $T_e$ at $r = 4 \, R_{\odot}$ and plots them
as a function of the wind speed $u_p$ in each model.
The general trend is for higher values of $T_p$ to be associated
with faster wind speeds, whereas $T_e$ does not vary as much.
Figure \ref{fig11}(b) shows hourly-averaged measurements of
$T_p$ and $T_e$ from {\em ISEE-3} at 1~AU \citep{Nb98}.
The similarities between the two trends are interesting, but
should be interpreted with caution.
Note that Figure \ref{fig11}(a) shows model results
consistent with one set of data over polar coronal holes,
but Figure \ref{fig11}(b) shows a wide swath of data taken over
several years in the ecliptic plane.
It is also true that measurements at intermediate distances
(e.g., {\em Helios} and {\em Parker Solar Probe}) show
a mild anticorrelation between $T_e$ and the wind speed;
see \citet{Mk20}.

\begin{figure}[!t]
\epsscale{1.17}
\plotone{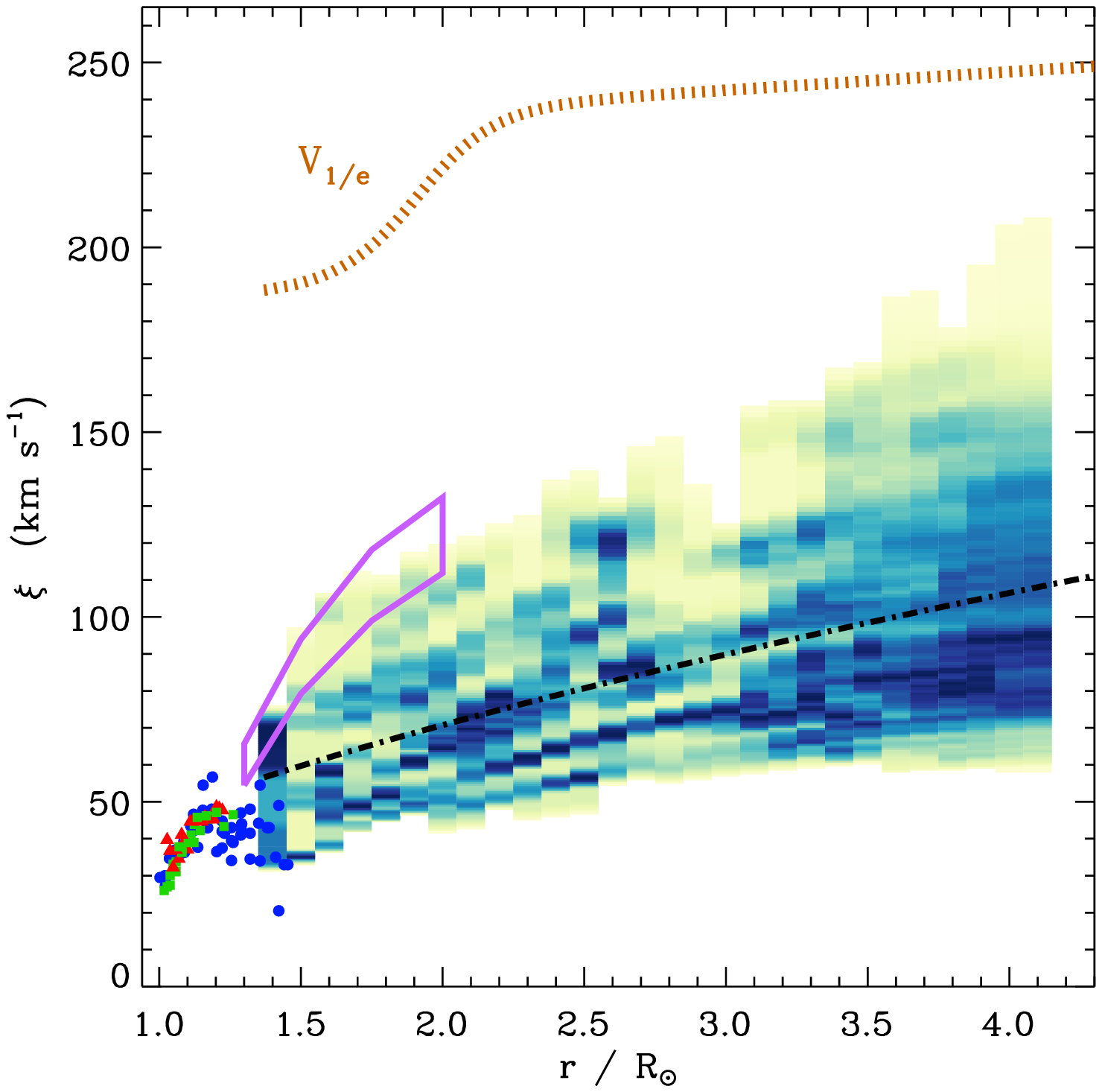}
\caption{Monte Carlo posterior distributions for the LOS-projected
turbulent nonthermal velocity $\xi$, shown using the same
color map as Figure \ref{fig08}.
Other observational results are duplicated from Figure \ref{fig04},
and the fitting function from Equation (\ref{eq:xifit}) is shown with
a black dot-dashed curve.
\label{fig12}}
\end{figure}

Figure~\ref{fig12} shows the distribution of turbulent velocity
amplitudes $\xi(r)$ for all 3,507 validated models.
The most probable values appear to fall {\em in between} the high
values inferred by \citet{Es99} and the low values inferred by
\citet{Hh13} and others.
If the various modeling assumptions described in Section \ref{sec:emp}
are to be believed, the UVCS data can provide a resolution to
the existing tension between these two seemingly incompatible
measurements.
Figure~\ref{fig12} also shows a relatively simple power-law fit to
the radial dependence of the median values of $\xi(r)$,
\begin{equation}
  \xi \, \approx \, 47 \, \left( \frac{r}{R_{\odot}} \right)^{0.59}
  \,\,\,\, \mbox{km s$^{-1}$}
  \label{eq:xifit}
\end{equation}
which was used in Equation (\ref{eq:wpperp}) to obtain the ion
temperatures shown in Figure \ref{fig09}.

Table~\ref{tab02} provides a summary of various statistical
properties of the posterior distributions at a selection of six
observation heights.
The values provided are for the median (50\% percentile) and $\pm$1
standard deviations away from the median (16\% and 84\% percentile)
for the distributions of $n_e$, $u_p$, $\xi$,
$T_e$, $T_{p \perp}$, and $T_{p \parallel}$.
Also shown are derived quantities for plasma heating rates that
are described in the Appendix.

\begin{table*}[!t]
\noindent
\small
\begin{center}
\caption{Plasma Parameters Consistent with UVCS Data}
\label{tab02}
\begin{tabular}{lcrrrrrr}
\hline
\hline
Quantity & Percentile &
$r = 1.5 \, R_{\odot}$ &
$r = 2.0 \, R_{\odot}$ &
$r = 2.5 \, R_{\odot}$ &
$r = 3.0 \, R_{\odot}$ &
$r = 3.5 \, R_{\odot}$ &
$r = 4.0 \, R_{\odot}$ \\
\hline
$n_e$ ($10^5$ cm$^{-3}$) & 
 16\%  &  11.7849 & 2.7427 & 1.0397 & 0.4490 & 0.2765 & 0.1732 \\
  &
 50\%  &  14.0308 & 3.4586 & 1.2988 & 0.5583 & 0.3841 & 0.2358 \\
  &
 84\%  &  15.9363 & 4.1464 & 1.4971 & 0.7243 & 0.4304 & 0.2760 \\
 
\hline
$u_p$ (km s$^{-1}$) &
 16\%  &    61.58 & 123.17 & 177.01 & 173.32 & 233.69 & 262.32 \\
  &
 50\%  &    84.59 & 143.12 & 205.82 & 236.04 & 275.90 & 314.27 \\
  &
 84\%  &   104.94 & 195.78 & 260.60 & 314.60 & 343.73 & 408.55 \\
 
\hline
$\xi$ (km s$^{-1}$) &
 16\%  &    35.87 &  54.16 &  60.84 &  69.93 &  75.22 &  80.21 \\
  &
 50\%  &    54.82 &  71.76 &  92.21 &  84.41 & 100.45 & 107.02 \\
  &
 84\%  &    78.92 &  90.50 & 119.23 & 107.99 & 131.97 & 138.41 \\
 
\hline
$T_e$ ($10^6$ K) &
 16\%  &    1.059 &  1.075 &  0.853 &  0.952 &  0.984 &  0.831 \\
  &
 50\%  &    1.208 &  1.361 &  1.151 &  1.194 &  1.269 &  1.208 \\
  &
 84\%  &    1.352 &  1.449 &  1.343 &  1.447 &  1.562 &  1.635 \\
 
\hline
$T_{p \parallel}$ ($10^6$ K) &
 16\%  &    0.844 &  0.931 &  1.362 &  1.285 &  1.433 &  0.632 \\
  &
 50\%  &    1.154 &  1.380 &  1.781 &  2.113 &  2.087 &  1.653 \\
  &
 84\%  &    1.645 &  2.986 &  3.526 &  4.037 &  3.481 &  3.775 \\
 
\hline
$T_{p \perp}$ ($10^6$ K) &
 16\%  &    1.161 &  1.124 &  1.271 &  1.259 &  1.119 &  1.193 \\
  &
 50\%  &    1.439 &  1.545 &  1.738 &  2.135 &  1.885 &  2.021 \\
  &
 84\%  &    1.633 &  2.501 &  3.105 &  3.109 &  2.914 &  3.138 \\

\hline
$Q_{p \parallel}/Q_{\rm norm}$ &  ...  &
  55.829 & 10.865 &  4.124 &  1.936 &  0.884 &  0.331 \\

$Q_{p \perp}/Q_{\rm norm}$ &  ...  &
  90.143 & 26.097 & 10.632 &  5.178 &  2.789 &  1.585 \\

$Q_e/Q_{\rm norm}$ &  ...  &
 164.431 & 35.004 & 13.840 &  6.984 &  3.993 &  2.485 \\

$Q_{\rm damp}/Q_{\rm norm}$ &  ...  &
 450.898 & 97.950 & 32.962 & 14.237 &  7.163 &  4.012 \\

\hline
\end{tabular}
\end{center}
\end{table*}

Lastly, additional data from this paper have been published
to a third-party repository \citep{Cr20}.
These data include tabulated information for Figures \ref{fig01},
\ref{fig03}(a), \ref{fig06}(a), and \ref{fig13}, as well
as the statistical properties given in Table \ref{tab02} for
the full grid of 28 radial distances (i.e., between 1.4 and
4.1~$R_{\odot}$ in increments of 0.1~$R_{\odot}$).
The tabulated H~I Ly$\alpha$ data for Figure \ref{fig01}
also include the exact dates and times for each relevant UVCS
exposure, so that the original data can be reassembled and
reanalyzed from the {\em SOHO} Archive
\citep[see, e.g.,][]{SD97,Os10}.
The repository contains the full set of Monte Carlo input and
output parameters for all 3,507 validated H~I Ly$\alpha$ profile models.

\section{Discussion and Conclusions}
\label{sec:conc}

Measurements of coronal temperatures, flow speeds, wave amplitudes,
and heating rates are necessary to have when testing (i.e.,
validating or falsifying) theoretical models.
A primary goal of this paper was to put new observational constraints
on the values of relevant proton and electron properties in
the acceleration region of the fast solar wind.
To accomplish this, a large Monte Carlo ensemble of trial models
of the corona was created, and H~I Ly$\alpha$ emission-line profiles
were simulated for each randomly generated model.
Part of this model construction procedure was to ensure
compatibility with mass and momentum conservation along open
field lines, \citep[see, e.g.,][]{LS16}.
Only about one out of every 800 of these models ended up agreeing
with the UVCS observations within an uncertainty range of $\pm 1$
standard deviations around measured intensities and line widths.
This subset of successful model parameters led to the construction
of posterior probability distributions for quantities such as
the wind speed, electron density, electron temperature,
proton temperature (parallel and perpendicular to the magnetic
field), and Alfv\'{e}n-wave amplitude.

Measured values of the outflow speed for the coupled proton--electron
plasma (shown in Figure \ref{fig08}) agree well with earlier results
from \citet{Cr99}, who analyzed the same UVCS data with a completely
different analysis procedure.
Measured temperatures were obtained with the help of a two-fluid
momentum conservation equation, such that there were separate
constraints on the thermal and nonthermal components of the
H~I Ly$\alpha$ line width.
Thermal temperatures did not show substantial radial dependence,
and typical values were approximately $T_{p \parallel} \approx 1.8$~MK,
$T_{p \perp} \approx 1.9$~MK, and $T_{e} \approx 1.2$~MK
(see Figures \ref{fig09}--\ref{fig10}).
Results for nonthermal line broadening---which were obtained in tandem
with a specific model of MHD turbulence---were consistent
with moderately damped LOS velocity amplitudes that increase from
about 50 km~s$^{-1}$ at 1.5~$R_{\odot}$ to about
100 km~s$^{-1}$ at 4~$R_{\odot}$ (see Figure \ref{fig12}).
These values fall {\em in between} two sets of seemingly incompatible
measurements of undamped \citep{Es99} and heavily damped \citep{Hh13}
Alfv\'{e}n waves in coronal holes.

The results presented above are plausible, but they should be
tested further with other analysis methods and observational data.
A larger number of Monte Carlo trials---possibly extending into
a larger volume of parameter space---would be useful as an improvement
on the statistics of the posterior distributions.
If possible, the use of mass and momentum conservation equations 
should be replaced by independent measurements of, say,
the solar wind outflow speed \citep[e.g.,][]{Df18}
and the electron temperature \citep[e.g.,][]{Rg11}.
The model of turbulent dissipation described in
Section \ref{sec:emp:Brxi} is only one possibility out of many
\citep[see also][]{Us18,Zk18,Ch19,Ma19}.
Also, the idea that nonthermal line broadening could be the result
of other features such as impulsive jets, short-lived filaments,
or magnetic ``switchbacks'' \citep[e.g.,][]{Tn20} should be explored.

Lastly, it would be advantageous to repeat this
analysis procedure with data from other source-regions of the
solar wind, such as equatorial coronal holes \citep{Mir01},
helmet streamers \citep{St02}, and pseudostreamers \citep{Wa12}.
Repeating the UVCS measurements with next-generation
coronagraph/spectrometer instruments \citep{Ko08,Vr18,Lam19}
would also likely provide new constraints on theoretical models.
In addition to using the H~I Ly$\alpha$ line, \citet{VC16} explored
how the inclusion of other hydrogen lines can provide complementary
information about the accelerating solar wind.
The {\em Metis} coronagraph on board {\em Solar Orbiter}
\citep{Rom17,An19} is poised to begin measurements of H~I Ly$\alpha$
Doppler dimming when its commissioning phase is completed.
A future instrument with greater sensitivity and a broader spectral
range than UVCS could measure the kinetic properties of dozens
of other heavy ions, and also may be able to detect subtle departures
from Gaussian line shapes that indicate the presence of specific
non-Maxwellian velocity distributions \citep[e.g.,][]{Cr01,Jf18}.

\vspace*{0.12in}
\hspace*{0.75in}ACKNOWLEDGMENTS
\vspace*{0.05in}

The author gratefully acknowledges
John Kohl, Leonard Strachan, Larry Gardner,
Mari Paz Miralles, Chris Gilly, Silvano Fineschi, Phil Isenberg,
Lika Guhathakurta, and Bill Matthaeus for many valuable discussions.
The author also thanks Sasha Panasyuk for key assistance in
performing the original H~I Ly$\alpha$ data reduction
and line-profile fitting.
The author is grateful to the anonymous referee for many
constructive suggestions that have improved this paper.
The author would also like to thank David M.\  Stern for creating
the Interactive Data Language (IDL), which receives far too little
praise despite being a critical building block of modern heliophysics.
This work was supported by the National Aeronautics and Space
Administration (NASA) under grants {NNX\-15\-AW33G} and
{NNX\-16\-AG87G}, and by the National Science Foundation (NSF)
under grants 1540094 and 1613207.
This research made extensive use of NASA's Astrophysics Data System (ADS).
The author acknowledges use of OMNI data from NASA's Space Physics
Data Facility OMNIWeb service.
UVCS is a joint program of the Smithsonian Astrophysical Observatory,
Agenzia Spaz\-i\-ale Italiana, and the Swiss contribution to the
ESA PRODEX program.
{\em SOHO} is a project of international cooperation between ESA and NASA.

\facilities{SOHO}

\appendix

\section{Preliminary Constraints on Heating Rates}
\label{sec:QpQe}

The main results of this paper for proton and electron
temperatures in coronal holes
(i.e., $T_p \approx T_e$ and $T_{p \perp} \approx T_{p \parallel}$)
seem rather ``mild.''
Other recent observational and theoretical studies point to the
likelihood of strong kinetic heating mechanisms---such as ion
cyclotron resonance and associated velocity-space diffusion---that
could provide $T_p \gg T_e$ and $T_{p \perp} \gg T_{p \parallel}$
\citep[e.g.,][]{HI02,Ma06,CPI10,Ka13,Cr14}.
However, the current observational results do not necessarily
rule out the presence of these effects.
Rather than examining local temperatures, it would be preferable to
know the actual plasma heating rates (i.e., in units of power
deposited per unit volume).
Unfortunately, computing these rates from the UVCS measurements
involves taking one additional step farther away from the
theory-agnostic ``empirical modeling'' approach described in
Section \ref{sec:emp}.

We proceed anyway by assuming the protons and electrons obey
time-steady conservation equations for thermal energy
\citep[see, e.g.,][]{Is84,LX99,CFK}.
The goal is to use the results of the Monte Carlo modeling exercise
to specify every term in these equations except the rates of
direct heat deposition, then solve for the latter.
The adopted electron energy equation assumes an isotropic velocity
distribution function that is affected by plasma expansion,
\citet{SH53} heat conduction, collisional equilibration with protons,
and some amount of direct heating $Q_e$, with
\begin{displaymath}
  \frac{3}{2} n_e u_p k_{\rm B} \frac{\partial T_e}{\partial r}
  - u_p k_{\rm B} T_e \frac{\partial n_e}{\partial r} \, = \,
\end{displaymath}
\begin{displaymath}
  Q_{e} - \frac{1}{A} \frac{\partial}{\partial r} ( A q_e )
  + C_{\parallel ep} (T_{p \parallel} - T_e)
\end{displaymath}
\begin{equation}
  + \, C_{\perp ep} (T_{p \perp} - T_e)
  + m_p ( J_{\parallel ep} + J_{\perp ep} )
  \label{eq:thermalQe}
\end{equation}
and the classical \citep{SH53} thermal conduction flux is given by
\begin{equation}
  q_e \, = \, - (1.84 \times 10^{-5} \,\, \mbox{erg}
  \,\, \mbox{cm}^{-1} \, \mbox{s}^{-1} \, \mbox{K}^{-7/2})
  \, \frac{T_e^{5/2}}{\ln \Lambda_{ee}}
  \, \frac{\partial T_e}{\partial r}
\end{equation}
(see, however, \citeauthor{SC20} \citeyear{SC20} for discussions
of nonclassical heat conduction in the extended corona) with the
electron Coulomb logarithm given approximately by
\begin{equation}
  \ln \Lambda_{ee} \, = \, 23.2 + \frac{3}{2} \ln \left(
  \frac{T_e}{10^{6} \, \mbox{K}} \right) - \frac{1}{2} \ln \left(
  \frac{n_e}{10^{6} \, \mbox{cm}^{-3}} \right)  \,\, .
\end{equation}
The Coulomb collision rates denoted by $C$ and $J$, with various
subscripts, are given in detail by \citet{Is84} and \citet{CFK}.
Proton energy conservation is expressed using two equations for the
bi-Maxwellian components $T_{p \parallel}$ and $T_{p \perp}$, with
\begin{displaymath}
  \frac{1}{2} n_p u_p k_{\rm B} \frac{\partial T_{p \parallel}}{\partial r}
  + n_p k_{\rm B} T_{p \parallel} \frac{\partial u_p}{\partial r}
  \, = \,
\end{displaymath}
\begin{equation}
  Q_{p \parallel} + C_{\parallel pe} (T_e - T_{p \parallel})
  + m_e J_{\parallel pe} \,\, ,
  \label{eq:thermalQppara}
\end{equation}
\begin{displaymath}
  n_p u_p k_{\rm B} \frac{\partial T_{p \perp}}{\partial r}
  + \frac{n_p k_{\rm B} T_{p \perp}}{A} \frac{\partial A}{\partial r}
  \, = \,
\end{displaymath}
\begin{equation}
  Q_{p \perp} + C_{\perp pe} (T_e - T_{p \perp})
  + m_e J_{\perp pe}
  \label{eq:thermalQpperp}
\end{equation}
and proton thermal conduction is neglected because it is traditionally
at least 20 times weaker than electron heat conduction
\citep[see, however,][]{Scud15}.

To evaluate many of the terms with radial derivatives, one specific
set of the original Monte Carlo parameters was chosen to produce
plasma parameters that pass through the median values given in
Table~\ref{tab02}.
To constrain the density and outflow speed, we used
$\Delta_{\rm lo} = 1.40$, $\Delta_{\rm hi} = 3.25$, 
$f_{\rm max} = 7.0$, and $u_{\infty} = 460$ km~s$^{-1}$.
For the temperatures, it was simpler to produce analytic
least-squares fits to the median model values.
Using $x = r / R_{\odot}$ as above,
\begin{equation}
  T_e \, = \, 1.31 x^{-0.05} \, \mbox{MK}
\end{equation}
\begin{equation}
  T_{p \parallel} \, = \,
  \left( -0.8177 + 1.5821 x - 0.2249 x^2 \right) \, \mbox{MK}
\end{equation}
\begin{equation}
  T_{p \perp} \, = \,
  \left( 0.3432 + 0.8125 x - 0.1001 x^2 \right) \, \mbox{MK}
\end{equation}
and these functions are valid only for $1.4 \leq x \leq 4.1$.
The turbulence amplitude $\xi$ (or $v_{\perp}$) was not needed
directly for evaluating the heating rates, but for later comparison
we took the above set of model parameters and computed a
representative curve for $Q_{\rm damp}$ from
Equations (\ref{eq:action}), (\ref{eq:xifit}),
and the turbulence dissipation model of \citet{CvB12}.

\begin{figure}[!t]
\epsscale{0.67}
\plotone{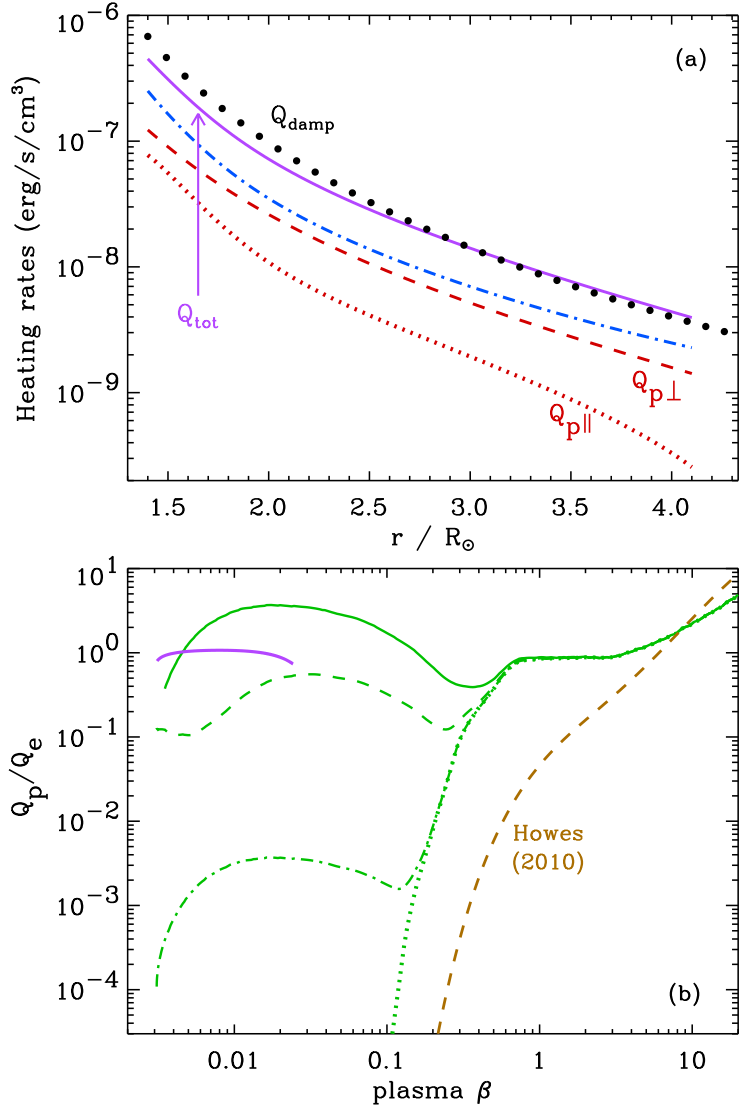}
\caption{Modeled heating rates $Q_e$ (blue dot-dashed curve),
$Q_{p \parallel}$ (red dotted curve),
$Q_{p \perp}$ (red dashed curve),
$Q_{\rm tot}$ (purple solid curve), and
$Q_{\rm damp}$ (black circles).
(b) UVCS-derived model result for $Q_p/Q_e$ (purple solid curve)
compared with theoretical heating ratios from
\citet{Hw10} (brown dashed curve) and
\citet{CvB12} (green curves).
The cascade/mode-coupling parameters for the latter are:
$F=1$, $\Phi = 10$ (solid green), 
$F=1$, $\Phi = 1$ (dashed green), 
$F=0.1$, $\Phi = 10$ (dot-dashed green),
$F=0.01$, $\Phi = 10$ (dotted green),
where $F$ is the number multiplied by the fast-mode energy
density $U_{\rm F}$ in Figure 16(b) of \citet{CvB12}.
\label{fig13}}
\end{figure}

Equations (\ref{eq:thermalQe})--(\ref{eq:thermalQpperp})
were thus solved for $Q_e$, $Q_{p \parallel}$, and $Q_{p \perp}$.
Figure~\ref{fig13}(a) shows these results as a function of
heliocentric distance.
Below, we also examine the total rate of proton heating
($Q_p = Q_{p \parallel} + Q_{p \perp}$) and the total rate of heating
for both species ($Q_{\rm tot} = Q_p + Q_e$).
Note that the proton and electron heating rates are roughly
comparable to one another in magnitude, with the ratio $Q_p/Q_e$
ranging between 0.73 and 1.07 over the modeled heights.
However, the heating anisotropy ratio $Q_{p \perp}/Q_{p \parallel}$
is consistently greater than unity, increasing from about 1.6
at low heights to 5.5 at large heights.
This preferential perpendicular heating does not show up as clearly
in $T_{p \parallel}$ and $T_{p \perp}$ because of the other heating
and cooling terms in
Equations (\ref{eq:thermalQppara})--(\ref{eq:thermalQpperp})
due to plasma expansion and Coulomb collisions.

Figure~\ref{fig13}(a) also compares $Q_{\rm tot}$ with the
wave/turbulence dissipation rate $Q_{\rm damp}$ that was computed
for the best-fitting model parameters described above.
The fact that these two rates agree so well with one another
(i.e., never differing by more than $\sim$30\%) was by no means
a foregone conclusion.
There could have been three possible outcomes:
\begin{enumerate}
\item
If $Q_{\rm damp} \gg Q_{\rm tot}$, this would point to a possible
inconsistency in the modeling assumptions.
In other words, if the energy lost by turbulent dissipation just
disappeared, and did not show up as plasma heating, then where
could it have gone?
It is possible this energy could be converted into other forms,
such as shocks or MHD waves (in modes other than the dominant
Alfv\'{e}n mode) or as bulk-flow kinetic energy in, e.g.,
reconnection exhausts.
However, those other forms often undergo rapid braking and
thermalization \citep[e.g.,][]{Bi09,Zh19}, so at least some of it
ought to appear in $Q_{\rm tot}$ eventually.
\item
If $Q_{\rm damp} \ll Q_{\rm tot}$, this would have implied that the
damping of MHD turbulence is insufficient to heat polar coronal holes.
If this was the case, then some other physical process,
such as slow nanoflare-type energy release \citep{SU81,P88},
would have to be responsible for supplying $Q_{\rm tot}$ in
these regions.
\item
If $Q_{\rm damp} \approx Q_{\rm tot}$, as we indeed see,
it implies that the model of MHD wave/turbulence dissipation described
in Section \ref{sec:emp:Brxi} appears to be sufficient to heat the
protons and electrons in coronal holes.
\end{enumerate}

Despite the fact that the UVCS data imply $Q_p \approx Q_e$, this
may still be consistent with protons and other ions being heated
by cyclotron resonance.
Figure \ref{fig13}(b) shows the ratio $Q_p/Q_e$ plotted versus
the plasma $\beta$ ratio for the model coronal hole.
Here, $\beta = (c_s / V_{\rm A})^2$, where $c_s$ is an effective
sound speed defined using the average one-fluid temperature
$(T_p + T_e)/2$.
It has been known for some time that the MHD turbulence excited
in low-$\beta$ plasmas does not naturally produce ion cyclotron
waves.
It tends to produce waves with low frequencies and high perpendicular
wavenumbers, which undergo linear Landau damping.
\citet{Hw10} predicted that in low-$\beta$ plasmas this would
give rise to $Q_p / Q_e \ll 1$, as shown in Figure \ref{fig13}(b).
However, \citet{CvB12} found that a small amount of nonlinear
mode-coupling between Alfv\'{e}n and fast-mode turbulence may produce
ion cyclotron waves that would heat protons when they dissipate.
The green curves in Figure \ref{fig13}(b) show that, depending on
the choice of mode-coupling parameters, it may be possible to
have $Q_p/Q_e \approx 1$ as indicated by the UVCS data.
Of course, several other mechanisms have been suggested to explain
preferential proton and ion heating in low-$\beta$ plasmas
\citep[e.g.,][]{VG04,Ch10,Sv15,Mt16,IV19,Sk19,Kw20}
and these should also be tested for consistency with the UVCS data.

Table~\ref{tab02} lists $Q_e$, $Q_{p \parallel}$, $Q_{p \perp}$,
and $Q_{\rm damp}$ at a selection of six observation heights.
These values are all divided by a constant
$Q_{\rm norm} = 10^{-9}$ erg~s$^{-1}$~cm$^{-3}$.
It is also possible to extrapolate the heating rates to larger
heliocentric distances and compare with in~situ measurements.
Using only the data points above $r = 3 \, R_{\odot}$, the radial
dependence of $Q_{\rm tot}$ is well-fit by a power-law function
$\propto r^{-4.1}$, and $Q_{\rm damp}$ is well-fit by $r^{-4.3}$.
This gives dissipation and heating rates of order $10^{-15}$ to
$10^{-14}$ erg~s$^{-1}$~cm$^{-3}$ at {\em Helios} distances
of order 0.3--0.6 AU, and values of a few times
$10^{-16}$ erg~s$^{-1}$~cm$^{-3}$ at 1~AU.
These values are comparable to empirically determined heating
rates from, e.g., \citet{Vs07} and \citet{Stv15}.


\begin{thebibliography}{}

\bibitem[Abbo et al.(2016)]{Ab16}
Abbo, L., Ofman, L., Antiochos, S. K., et al. 2016, \ssr, 201, 55

\bibitem[Aellig et al.(1998)]{Ae98}
Aellig, M. R., Gr\"{u}nwaldt, H., Bochsler, P., et al. 1998,
\jgr, 103, 17215

\bibitem[Alazraki \& Couturier(1971)]{AC71}
Alazraki, G., \& Couturier, P. 1971, \aap, 13, 380

\bibitem[Alfv\'{e}n(1941)]{A41}
Alfv\'{e}n, H. 1941, Arkiv f\"{o}r Matematik, Astronomi och Fysik
(Band 27A), 25, 1

\bibitem[Allen et al.(2000)]{Al00}
Allen, L. A., Habbal, S. R., \& Li, X. 2000, \jgr, 105, 23123

\bibitem[Antonucci(2006)]{An06}
Antonucci, E. 2006, \ssr, 124, 35

\bibitem[Antonucci et al.(2000)]{An00}
Antonucci, E., Dodero, M. A., \& Giordano, S. 2000, \solphys, 197, 115

\bibitem[Antonucci et al.(2019)]{An19}
Antonucci, E., Romoli, M., Andretta, V., et al. 2019,
\aap, in press, arXiv:1911.08462

\bibitem[Aschwanden \& Acton(2001)]{AA01}
Aschwanden, M. J., \& Acton, L. W. 2001, \apj, 550, 475

\bibitem[Banaszkiewicz et al.(1998)]{Ba98}
Banaszkiewicz, M., Axford, W. I., \& McKenzie, J. F. 1998,
\aap, 337, 940

\bibitem[Banerjee et al.(1998)]{Bj98}
Banerjee, D., Teriaca, L., Doyle, J. G., \& Wilhelm, K. 1998,
\aap, 339, 208

\bibitem[Beckers \& Chipman(1974)]{BC74}
Beckers, J. M., \& Chipman, E. 1974, \solphys, 34, 151

\bibitem[Belcher(1971)]{B71}
Belcher, J. W. 1971, \apj, 168, 509

\bibitem[Bemporad(2017)]{Bm17}
Bemporad, A. 2017, \apj, 846, 86

\bibitem[Bemporad \& Abbo(2012)]{BA12}
Bemporad, A., \& Abbo, L. 2012, \apj, 751, 110

\bibitem[Bertaux et al.(1997)]{Bx97}
Bertaux, J. L., Qu\'{e}merais, E., Lallement, R., et al. 1997,
\solphys, 175, 737

\bibitem[Birn et al.(2009)]{Bi09}
Birn, J., Fletcher, L., Hesse, M., \& Neukirch, T. 2009, \apj, 695, 1151

\bibitem[Bretherton \& Garrett(1968)]{BG68}
Bretherton, F. P., \& Garrett, C. J. R. 1968,
Proc.\  Roy.\  Soc.\  A, 302, 529

\bibitem[Chandran(2010)]{Ch10}
Chandran, B. D. G. 2010, \apj, 720, 548

\bibitem[Chandran \& Perez(2019)]{Ch19}
Chandran, B. D. G., \& Perez, J. C. 2019, J.\  Plasma Phys.,
85, 905850409

\bibitem[Chandran et al.(2010)]{CPI10}
Chandran, B. D. G., Pongkitiwanichakul, P., Isenberg, P. A., et al.
2010, \apj, 722, 710

\bibitem[Cram(1976)]{Cram76}
Cram, L. E. 1976, \solphys, 48, 3

\bibitem[Cranmer(1998)]{Cr98}
Cranmer, S. R. 1998, \apj, 508, 925

\bibitem[Cranmer(2001)]{Cr01}
Cranmer, S. R. 2001, \jgr, 106, 24937

\bibitem[Cranmer(2009)]{Cr09}
Cranmer, S. R. 2009, Living Rev.\  Solar Phys., 6, 3

\bibitem[Cranmer(2014)]{Cr14}
Cranmer, S. R. 2014, \apjs, 213, 16

\bibitem[Cranmer(2020)]{Cr20}
Cranmer, S. R. 2020, Heating Rates for Protons and Electrons in Polar
Coronal Holes: Empirical Constraints from UVCS/SOHO (Associated Data),
v1.0, Zenodo, doi:10.5281/zenodo.3908519

\bibitem[Cranmer et al.(1999a)]{CFK}
Cranmer, S. R., Field, G. B., \& Kohl, J. L. 1999a, \apj, 518, 937

\bibitem[Cranmer et al.(2010)]{Cr10}
Cranmer, S. R., Gardner, L. D., \& Kohl, J. L. 2010,
\solphys, 263, 275

\bibitem[Cranmer et al.(2017)]{Cr17}
Cranmer, S. R., Gibson, S. E., \& Riley, P. 2017, \ssr, 212, 1345

\bibitem[Cranmer et al.(1999b)]{Cr99}
Cranmer, S. R., Kohl, J. L., Noci, G., et al. 1999b, \apj, 511, 481

\bibitem[Cranmer et al.(2008)]{Cr08}
Cranmer, S. R., Panasyuk, A. V., \& Kohl, J. L. 2008, \apj, 678, 1480

\bibitem[Cranmer \& van Ballegooijen(2005)]{CvB05}
Cranmer, S. R., \& van Ballegooijen, A. A. 2005, \apjs, 156, 265

\bibitem[Cranmer \& van Ballegooijen(2012)]{CvB12}
Cranmer, S. R., \& van Ballegooijen, A. A. 2012, \apj, 754, 92

\bibitem[Cranmer et al.(2007)]{CvB07}
Cranmer, S. R., van Ballegooijen, A. A., \& Edgar, R. J. 2007,
\apjs, 171, 520

\bibitem[Cranmer \& Winebarger(2019)]{CW19}
Cranmer, S. R., \& Winebarger, A. R. 2019, \araa, 57, 157

\bibitem[Curdt et al.(2001)]{Cu01}
Curdt, W., Brekke, P., Feldman, U., et al. 2001, \aap, 375, 591

\bibitem[DeForest et al.(2018)]{Df18}
DeForest, C. E., Howard, R. A., Velli, M., et al. 2018, \apj, 862, 18

\bibitem[Dere et al.(1997)]{Dr97}
Dere, K. P., Landi, E., Mason, H. E., et al. 1997, \aaps, 125, 149

\bibitem[Dolei et al.(2016)]{Do16}
Dolei, S., Spadaro, D., \& Ventura, R. 2016, \aap, 592, A137

\bibitem[Dolei et al.(2018)]{Do18}
Dolei, S., Susino, R., Sasso, C., et al. 2018, \aap, 612, A84

\bibitem[Domingo et al.(1995)]{Do95}
Domingo, V., Fleck. B., \& Poland, A. I. 1995, \solphys, 162, 1

\bibitem[Doschek et al.(2001)]{Ds01}
Doschek, G. A., Feldman, U., Laming, J. M., Sch\"{u}hle, U., \&
Wilhelm, K. 2001, \apj, 546, 559

\bibitem[Doyle et al.(1999)]{Doy99}
Doyle, J. G., Teriaca, L., \& Banerjee, D. 1999, \aap, 349, 956

\bibitem[Esser(1990)]{Es90}
Esser, R. 1990, \jgr, 95, 10261

\bibitem[Esser \& Edgar(2000)]{EE00}
Esser, R., \& Edgar, R. J. 2000, \apjl, 532, L71

\bibitem[Esser et al.(1999)]{Es99}
Esser, R., Fineschi, S., Dobrzycka, D., et al. 1999, \apjl, 510, L63

\bibitem[Fineschi et al.(1998)]{Fi98}
Fineschi, S., Gardner, L. D., Kohl, J. L., Romoli, M., \&
Noci, G. 1998, Proc.\  SPIE, 3443, 67

\bibitem[Fisher \& Guhathakurta(1995)]{FG95}
Fisher, R. R., \& Guhathakurta, M. 1995, \apjl, 447, L139

\bibitem[Fleck \& \v{S}vestka(1997)]{FS97}
Fleck, B., \& \v{S}vestka, Z., eds. 1997, The First Results
from SOHO (Dordrecht: Kluwer)

\bibitem[Fludra et al.(1999)]{Fd99}
Fludra, A., Del Zanna, G., Alexander, D., \& Bromage, B. J. I. 1999,
\jgr, 104, 9709

\bibitem[Foley et al.(1997)]{Fo97}
Foley, C. R., Culhane, J. L., \& Acton, L. W. 1997,
\apj, 491, 933

\bibitem[Gabriel(1971)]{Gb71}
Gabriel, A. H. 1971, \solphys, 21, 392

\bibitem[Galvin \& Kohl(1999)]{GK99}
Galvin, A. B., \& Kohl, J. L. 1999, \jgr, 104, 9673

\bibitem[Gardner et al.(1996)]{Ga96}
Gardner, L. D., Kohl, J. L., Daigneau, P. S., et al. 1996,
Proc.\  SPIE, 2831, 2

\bibitem[Gardner et al.(2002)]{Ga02}
Gardner, L. D., Smith, P. L., Kohl, J. L., et al. 2002,
in The Radiometric Calibration of SOHO, ed. A. Pauluhn, M. Huber,
\& R. von Steiger (ESA SR-002; Noordwijk: ESA), 161

\bibitem[Gilly \& Cranmer(2020)]{GC20}
Gilly, C. R., \& Cranmer, S. R. 2020, \apj, submitted

\bibitem[Gray(1973)]{Gr73}
Gray, D. F. 1973, \apj, 184, 461

\bibitem[Guhathakurta et al.(1999)]{Gu99}
Guhathakurta, M., Fludra, A., Gibson, S. E., et al. 1999,
\jgr, 104, 9801

\bibitem[Gupta(2017)]{Gu17}
Gupta, G. R. 2017, \apj, 836, 4

\bibitem[Hahn et al.(2012)]{Hh12}
Hahn, M., Landi, E., \& Savin, D. W. 2012, \apj, 753, 36

\bibitem[Hahn \& Savin(2013)]{Hh13}
Hahn, M., \& Savin, D. W. 2013, \apj, 776, 78

\bibitem[Harvey \& Sheeley(1979)]{HS79}
Harvey, J. W., \& Sheeley, N. R., Jr. 1979, \ssr, 23, 139

\bibitem[Heinemann \& Olbert(1980)]{HO80}
Heinemann, M., \& Olbert, S. 1980, \jgr, 85, 1311

\bibitem[Hollweg(1973)]{H73}
Hollweg, J. V. 1973, \apj, 181, 547

\bibitem[Hollweg \& Isenberg(2002)]{HI02}
Hollweg, J. V., \& Isenberg, P. A. 2002, \jgr, 107, 1147

\bibitem[Howes(2010)]{Hw10}
Howes, G. G. 2010, \mnras, 409, L104

\bibitem[Hughes(1965)]{Hu65}
Hughes, C. J. 1965, \apj, 142, 321

\bibitem[Hummer(1962)]{H62}
Hummer, D. 1962, \mnras, 125, 21

\bibitem[Inhester(2015)]{In15}
Inhester, B. 2015, preprint, arXiv:1512.00651

\bibitem[Isenberg(1984)]{Is84}
Isenberg, P. A. 1984, \jgr, 89, 6613

\bibitem[Isenberg \& Vasquez(2019)]{IV19}
Isenberg, P. A., \& Vasquez, B. J. 2019, \apj, 887, 63

\bibitem[Jeffrey et al.(2018)]{Jf18}
Jeffrey, N. L. S., Hahn, M., Savin, D. W., \& Fletcher, L. 2018,
\apjl, 855, L13

\bibitem[Jones et al.(1992)]{Jn92}
Jones, H. P., Duvall, T. L., Harvey, J. W., et al. 1992, \solphys,
139, 211

\bibitem[Kasper et al.(2013)]{Ka13}
Kasper, J. C., Maruca, B. A., Stevens, M. L., et al. 2013,
\prl, 110, 091102

\bibitem[Kasper et al.(2007)]{Ka07}
Kasper, J. C., Stevens, M. L., Lazarus, A. J., et al. 2007,
\apj, 660, 901

\bibitem[Kawazura et al.(2020)]{Kw20}
Kawazura, Y., Schekochihin, A. A., Barnes, M., et al. 2020,
preprint, arXiv:2004.04922

\bibitem[King \& Papitashvili(2005)]{KP05}
King, J. H., \& Papitashvili, N. E. 2005, \jgr, 110, A02104

\bibitem[Klimchuk(2015)]{Kl15}
Klimchuk, J. A. 2015, Phil.\  Trans.\  Roy.\  Soc.\  A,
373, 20140256

\bibitem[Ko et al.(1997)]{YKo97}
Ko, Y.-K., Fisk, L. A., Geiss, J., Gloeckler, G., \&
Guhathakurta, M. 1997, \solphys, 171, 345

\bibitem[Kohl et al.(1999)]{Ko99}
Kohl, J. L., Esser, R., Cranmer, S. R., et al. 1999, \apjl, 510, L59

\bibitem[Kohl et al.(1995)]{Ko95}
Kohl, J. L., Esser, R., Gardner, L. D., et al. 1995, \solphys, 162, 313

\bibitem[Kohl et al.(2008)]{Ko08}
Kohl, J. L., Jain, R., Cranmer, S. R., et al. 2008,
J.\  Ap.\  Astron., 29, 321

\bibitem[Kohl et al.(1997)]{Ko97}
Kohl, J. L., Noci, G., Antonucci, E., et al. 1997, \solphys, 175, 613

\bibitem[Kohl et al.(2006a)]{Ko06a}
Kohl, J. L., Noci, G., Cranmer, S. R., \& Raymond, J. C. 2006a,
\aapr, 13, 31

\bibitem[Kohl et al.(2006b)]{Ko06tail}
Kohl, J. L., Panasyuk, A. V., Cranmer, S. R., et al. 2006b,
in SOHO-17: Ten Years of SOHO and Beyond, ed. H. Lacoste \& L. Ouwehand
(ESA SP-617; Noordwijk: ESA), 25

\bibitem[Kohl et al.(1980)]{Ko80}
Kohl, J. L., Weiser, H., Withbroe, G. L., et al. 1980, \apjl, 241, L117

\bibitem[Kopp \& Holzer(1976)]{KH76}
Kopp, R. A., \& Holzer, T. E. 1976, \solphys, 49, 43

\bibitem[Laming et al.(2019)]{Lam19}
Laming, J. M., Vourlidas, A., Korendyke, C., et al. 2019,
\apj, 879, 124

\bibitem[Landi(2008)]{La08}
Landi, E. 2008, \apj, 685, 1270

\bibitem[Landi \& Cranmer(2009)]{LC09}
Landi, E., \& Cranmer, S. R. 2009, \apj, 691, 794

\bibitem[Landi et al.(2014)]{Ln14}
Landi, E., Oran, R., Lepri, S. T., et al. 2014, \apj, 790, 111

\bibitem[Landi et al.(2013)]{Ln13}
Landi, E., Young, P. R., Dere, K. P., et al. 2013, \apj, 763, 86

\bibitem[Lemaire \& Katsiyannis(2020)]{LK20}
Lemaire, J. F., \& Katsiyannis, A. C. 2020, \solphys, submitted,
arXiv:2002.07495

\bibitem[Lemaire \& Stegen(2016)]{LS16}
Lemaire, J. F., \& Stegen, K. 2016, \solphys, 291, 3659

\bibitem[Li et al.(1999)]{LX99}
Li, X., Habbal, S. R., Hollweg, J. V., et al. 1999, \jgr, 104, 2521

\bibitem[Linker et al.(1999)]{Li99}
Linker, J. A., Miki\'{c}, Z., Biesecker, D. A., et al. 1999,
\jgr, 104, 9809

\bibitem[Livingston et al.(1976)]{Li76}
Livingston, W. C., Harvey, J., Pierce, A. K., et al. 1976,
Applied Optics, 15, 33

\bibitem[Maksimovic et al.(2020)]{Mk20}
Maksimovic, M., Bale, S. D., Ber\v{c}i\v{c}, L., et al. 2020,
\apjs, 246, 62

\bibitem[Mallet et al.(2019)]{Ma19}
Mallet, A., Klein, K. G., Chandran, B. D. G., et al. 2019,
J.\  Plasma Phys., 85, 175850302

\bibitem[Maltby(1968)]{Ma68}
Maltby, P. 1968, \solphys, 5, 3

\bibitem[Marquardt(1963)]{Mq63}
Marquardt, D. W. 1963, J.\  Soc.\  Indust.\  Appl.\  Math., 11, 431

\bibitem[Marsch(2006)]{Ma06}
Marsch, E. 2006, Living Rev.\  Solar Phys., 3, 1

\bibitem[Matteini et al.(2007)]{Me07}
Matteini, L., Landi, S., Hellinger, P., et al. 2007, \grl, 34, L20105

\bibitem[Matthaeus et al.(2016)]{Mt16}
Matthaeus, W. H., Parashar, T. N., Wan, M., \& Wu, P. 2016,
\apjl, 827, L7

\bibitem[Miki\'{c} et al.(1999)]{Mk99}
Miki\'{c}, Z., Linker, J. A., Schnack, D. D., et al. 1999,
Phys.\  Plasmas, 6, 2217

\bibitem[Minnaert(1930)]{Min30}
Minnaert, M. 1930, \zap, 1, 209

\bibitem[Miralles et al.(2001)]{Mir01}
Miralles, M. P., Cranmer, S. R., Panasyuk, A. V., Romoli, M., \&
Kohl, J. L. 2001, \apjl, 549, L257

\bibitem[Munro \& Jackson(1977)]{MJ77}
Munro, R. H., \& Jackson, B. V. 1977, \apj, 213, 874

\bibitem[Nakagawa(2008)]{Nk08}
Nakagawa, A. 2008, \apj, 674, 1167

\bibitem[Newbury et al.(1998)]{Nb98}
Newbury, J. A., Russell, C. T., Phillips, J. L., et al. 1998,
\jgr, 103, 9553

\bibitem[Noci et al.(1997)]{No97}
Noci, G., Kohl, J. L., Antonucci, E., et al. 1997,
Adv.\  Space Res., 20, 2219

\bibitem[Noci \& Maccari(1999)]{NM99}
Noci, G., \& Maccari, L. 1999, \aap, 341, 275

\bibitem[Olsen et al.(1994)]{Ol94}
Olsen, E. L., Leer, E., \& Holzer, T. E. 1994, \apj, 420, 913

\bibitem[Osuna et al.(2010)]{Os10}
Osuna, P., Arviset, C., Baines, D., et al. 2010,
in ASP Conf.\  Proc.\  434, Astronomical Data Analysis Software and
Systems XIX, ed. Y. Mizumoto, K. Morita, \& M. Ohishi
(San Francisco, CA: ASP), 3

\bibitem[Owocki et al.(1983)]{Ow83}
Owocki, S. P., Holzer, T. E., \& Hundhausen, A. J. 1983, \apj, 275, 354

\bibitem[Parker(1965)]{P65}
Parker, E. N. 1965, \ssr, 4, 666

\bibitem[Parker(1988)]{P88}
Parker, E. N. 1988, \apj, 330, 474

\bibitem[Parnell \& De Moortel(2012)]{PD12}
Parnell, C. E., \& De Moortel, I. 2012,
Phil.\  Trans.\  Roy.\  Soc.\  A, 370, 3217

\bibitem[Peter(2015)]{Pe15}
Peter, H. 2015, Phil.\  Trans.\  Roy.\  Soc.\  A,
373, 20150055

\bibitem[Raymond et al.(1997)]{Ra97}
Raymond, J. C., Kohl, J. L., Noci, G., et al. 1997, \solphys, 175, 645

\bibitem[Reginald \& Davila(2000)]{Rg00}
Reginald, N. L., \& Davila, J. M. 2000, \solphys, 195, 111

\bibitem[Reginald et al.(2011)]{Rg11}
Reginald, N. L., Davila, J. M., St.~Cyr, O. C., et al. 2011,
\solphys, 270, 235

\bibitem[Riley(2007)]{Ri07}
Riley, P. 2007, J.\  Atmos.\  Sol.-Terr.\  Phys., 69, 32

\bibitem[Riley et al.(2001)]{Ri01}
Riley, P., Linker, J. A., \& Miki\'{c}, Z. 2001, \jgr, 106, 15889

\bibitem[Romoli et al.(2017)]{Rom17}
Romoli, M., Landini, F., Antonucci, E., et al. 2017,
Proc.\  SPIE, 10563, 105631M

\bibitem[Saito et al.(1970)]{Saito}
Saito, K., Makita, M., Nishi, K., \& Hata, S. 1970,
Ann.\  Tokyo Astron.\  Obs., 12, 53

\bibitem[Sanchez Duarte(1997)]{SD97}
Sanchez Duarte, L. 1997, in ASP Conf.\  Proc.\  118,
First Advances in Solar Physics Euroconference: Advances in Physics of
Sunspots, ed. B. Schmieder, J. del Toro Iniesta, \& M. Vazquez
(San Francisco, CA: ASP), 378

\bibitem[Schekochihin et al.(2019)]{Sk19}
Schekochihin, A. A., Kawazura, Y., \& Barnes, M. A. 2019,
J.\  Plasma Phys., 85, 905850303

\bibitem[Scherrer et al.(1995)]{Sc95}
Scherrer, P. H., Bogart, R. S., Bush, R. I., et al. 1995,
\solphys, 162, 129

\bibitem[Schiff \& Cranmer(2020)]{SC20}
Schiff, A. J., \& Cranmer, S. R. 2020, \apj, submitted

\bibitem[Scudder(2015)]{Scud15}
Scudder, J. D. 2015, \apj, 809, 126

\bibitem[Servidio et al.(2015)]{Sv15}
Servidio, S., Valentini, F., Perrone, D., et al. 2015,
J.\  Plasma Phys., 81, 325810107

\bibitem[Spadaro et al.(2017)]{Sp17}
Spadaro, D., Susino, R., Dolei, S., et al. 2017, \aap, 603, A35

\bibitem[Spitzer \& H\"{a}rm(1953)]{SH53}
Spitzer, L., \& H\"{a}rm, R. 1953, Phys.\   Rev., 89, 977

\bibitem[Strachan et al.(1993)]{St93}
Strachan, L., Kohl, J. L., Weiser, H., et al. 1993, \apj, 412, 410

\bibitem[Strachan et al.(2012)]{St12}
Strachan, L., Panasyuk, A. V., Kohl, J. L., et al. 2012, \apj, 745, 51

\bibitem[Strachan et al.(2002)]{St02}
Strachan, L., Suleiman, R., Panasyuk, A. V., Biesecker, D. A.,
\& Kohl, J. L. 2002, \apj, 571, 1008

\bibitem[Struve \& Elvey(1934)]{SE34}
Struve, O., \& Elvey, C. T. 1934, \apj, 79, 409

\bibitem[Sturrock \& Uchida(1981)]{SU81}
Sturrock, P. A., \& Uchida, Y. 1981, \apj, 246, 331

\bibitem[\v{S}tver\'{a}k et al.(2015)]{Stv15}
\v{S}tver\'{a}k, \v{S}., Tr\'{a}vn\'{\i}\v{c}ek, P. M.,
\& Hellinger, P. 2015, \jgr, 120, 8177

\bibitem[Suleiman et al.(1999)]{Su99}
Suleiman, R. M., Kohl, J. L., Panasyuk, A. V., et al. 1999,
\ssr, 87, 327

\bibitem[Tenerani et al.(2020)]{Tn20}
Tenerani, A., Velli, M., Matteini, L., et al. 2020, \apjs, 246, 32

\bibitem[Usmanov et al.(2018)]{Us18}
Usmanov, A. V., Matthaeus, W. H., Goldstein, M. L., \& Chhiber, R.
2018, \apj, 865, 25

\bibitem[van Ballegooijen \& Asgari-Targhi(2016)]{vB16}
van Ballegooijen, A. A., \& Asgari-Targhi, M. 2016, \apj, 821, 106

\bibitem[van de Hulst(1950)]{vd50}
van de Hulst, H. C. 1950, Bull.\  Astron.\  Inst.\  Netherlands, 11, 135

\bibitem[Vasquez et al.(2007)]{Vs07}
Vasquez, B. J., Smith, C. W., Hamilton, K., et al. 2007,
\jgr, 112, A07101

\bibitem[Verscharen et al.(2019)]{Ve19}
Verscharen, D., Klein, K. G., \& Maruca, B. A. 2019,
Living Rev.\  Solar Phys., 16, 5

\bibitem[Vial \& Chane-Yook(2016)]{VC16}
Vial, J.-C., \& Chane-Yook, M. 2016, \solphys, 291, 3549

\bibitem[Voitenko \& Goossens(2004)]{VG04}
Voitenko, Y., \& Goossens, M. 2004, \apjl, 605, L149

\bibitem[Vourlidas et al.(2018)]{Vr18}
Vourlidas, A., Ko, Y.-K., Laming, J. M., et al. 2018,
Eos Trans.\  AGU, Fall Meet.\  Suppl., abstract SH34A-02

\bibitem[Wang(2009)]{Wa09}
Wang, Y.-M. 2009, \ssr, 144, 383

\bibitem[Wang et al.(2012)]{Wa12}
Wang, Y. -M., Grappin, R., Robbrecht, E., \& Sheeley, N. R., Jr.
2012, \apj, 749, 182

\bibitem[Wilhelm(2006)]{Wm06}
Wilhelm, K. 2006, \aap, 455, 697

\bibitem[Withbroe et al.(1982)]{Wi82}
Withbroe, G. L., Kohl, J. L., Weiser, H., \& Munro, R. H. 1982,
\ssr, 33, 17

\bibitem[Woods et al.(2000)]{Wo00}
Woods, T. N., Tobiska, W. K., Rottman, G. J., \& Worden, J. R. 2000,
\jgr, 105, 27195

\bibitem[Woolsey \& Cranmer(2015)]{WC15}
Woolsey, L. N., \& Cranmer, S. R. 2015, \apj, 811, 136

\bibitem[Zangrilli et al.(1999)]{Zn99}
Zangrilli, L., Nicolosi, P., Poletto, G., et al. 1999, \aap, 342, 592

\bibitem[Zangrilli et al.(2002)]{Zn02}
Zangrilli, L., Poletto, G., Nicolosi, P., et al. 2002, \apj, 574, 477

\bibitem[Zank et al.(2018)]{Zk18}
Zank, G. P., Adhikari, L., Hunana, P., et al. 2018, \apj, 854, 32

\bibitem[Zhang et al.(2019)]{Zh19}
Zhang, Q., Drake, J. F., \& Swisdak, M. 2019, Phys.\  Plasmas,
26, 072115

\end{thebibliography}
\end{document}